\documentclass[12pt,prb,preprint,superscriptaddress,floatfix]{revtex4-1}

\usepackage{amsfonts,amssymb,amsthm,amsmath,latexsym}
\usepackage{subeqnarray,indent,url,bm}
\usepackage{graphicx}

\date{February 26, 2020; revised October 30, 2020 and May 15, 2021;
       final revision April 5, 2022 \\[1cm]}

\def\singlespace{ \renewcommand{\baselinestretch}{1} \large\normalsize }

\begin{document}

\title{\quad Skier and loop-the-loop with friction}

\author{Dominik Kufel}
\email{dominik.kufel.17@ucl.ac.uk (Third-year undergraduate student
   at the time of writing)}
\affiliation{\hbox{Department of Physics and Astronomy,
                   University College London, London WC1E 6BT, UK}}
\author{Alan D.~Sokal}
\email{sokal@nyu.edu}
\affiliation{Department of Mathematics, University College London,
                    London WC1E 6BT, UK}
\affiliation{Department of Physics, New York University,
                    New York, NY 10003, USA}


\singlespace 
\begin{abstract}
We solve analytically the differential equations for
a skier on a hemispherical hill
and for a particle on a loop-the-loop track
when the hill or track is endowed with
a coefficient of kinetic friction $\mu$.
For each problem, we determine the exact ``phase diagram''
in the two-dimensional parameter plane.
\bigskip
\begin{center}
\large To be published in the {\em American Journal of Physics}
\end{center}
\end{abstract}


\maketitle
\thispagestyle{empty}   

\clearpage

\newtheorem{theorem}{Theorem}[section]
\newtheorem{proposition}[theorem]{Proposition}
\newtheorem{lemma}[theorem]{Lemma}
\newtheorem{corollary}[theorem]{Corollary}
\newtheorem{definition}[theorem]{Definition}
\newtheorem{conjecture}[theorem]{Conjecture}
\newtheorem{question}[theorem]{Question}
\newtheorem{problem}[theorem]{Problem}
\newtheorem{example}[theorem]{Example}

\renewcommand{\theenumi}{\alph{enumi}}
\renewcommand{\labelenumi}{(\theenumi)}
\def\eop{\hbox{\kern1pt\vrule height6pt width4pt
depth1pt\kern1pt}\medskip}
\def\prf{\par\noindent{\bf Proof.\enspace}\rm}
\def\rmk{\par\medskip\noindent{\bf Remark\enspace}\rm}

\newcommand{\textbfit}[1]{\textbf{\textit{#1}}}

\newcommand{\bigdash}{%
\smallskip\begin{center} \rule{5cm}{0.1mm} \end{center}\smallskip}

\newcommand{\safepar}{ {\protect\hfill\protect\break\hspace*{5mm}} }

\newcommand{\be}{\begin{equation}}
\newcommand{\ee}{\end{equation}}
\newcommand{\<}{\langle}
\renewcommand{\>}{\rangle}
\newcommand{\widebar}{\overline}
\def\reff#1{(\protect\ref{#1})}
\def\spose#1{\hbox to 0pt{#1\hss}}
\def\ltapprox{\mathrel{\spose{\lower 3pt\hbox{$\mathchar"218$}}
    \raise 2.0pt\hbox{$\mathchar"13C$}}}
\def\gtapprox{\mathrel{\spose{\lower 3pt\hbox{$\mathchar"218$}}
    \raise 2.0pt\hbox{$\mathchar"13E$}}}
\def\textprime{${}^\prime$}
\def\proof{\par\medskip\noindent{\sc Proof.\ }}
\def\firstproof{\par\medskip\noindent{\sc First Proof.\ }}
\def\secondproof{\par\medskip\noindent{\sc Second Proof.\ }}
\def\alternateproof{\par\medskip\noindent{\sc Alternate Proof.\ }}
\def\algebraicproof{\par\medskip\noindent{\sc Algebraic Proof.\ }}
\def\combinatorialproof{\par\medskip\noindent{\sc Combinatorial Proof.\ }}
\def\proofof#1{\bigskip\noindent{\sc Proof of #1.\ }}
\def\firstproofof#1{\bigskip\noindent{\sc First Proof of #1.\ }}
\def\secondproofof#1{\bigskip\noindent{\sc Second Proof of #1.\ }}
\def\thirdproofof#1{\bigskip\noindent{\sc Third Proof of #1.\ }}
\def\algebraicproofof#1{\bigskip\noindent{\sc Algebraic Proof of #1.\ }}
\def\combinatorialproofof#1{\bigskip\noindent{\sc Combinatorial Proof of #1.\ }}
\def\sketchofproof{\par\medskip\noindent{\sc Sketch of proof.\ }}
\renewcommand{\qed}{ $\square$ \bigskip}
\newcommand{\myendremark}{ $\blacksquare$ \bigskip}
\def\half{ {1 \over 2} }
\def\third{ {1 \over 3} }
\def\twothird{ {2 \over 3} }
\def\smfrac#1#2{{\textstyle{#1\over #2}}}
\def\smhalf{ {\smfrac{1}{2}} }
\def\smquarter{ {\smfrac{1}{4}} }
\newcommand{\real}{\mathop{\rm Re}\nolimits}
\renewcommand{\Re}{\mathop{\rm Re}\nolimits}
\newcommand{\imag}{\mathop{\rm Im}\nolimits}
\renewcommand{\Im}{\mathop{\rm Im}\nolimits}
\newcommand{\sgn}{\mathop{\rm sgn}\nolimits}
\newcommand{\tr}{\mathop{\rm tr}\nolimits}
\newcommand{\tg}{\mathop{\rm tg}\nolimits}
\newcommand{\supp}{\mathop{\rm supp}\nolimits}
\newcommand{\disc}{\mathop{\rm disc}\nolimits}
\newcommand{\diag}{\mathop{\rm diag}\nolimits}
\newcommand{\csch}{\mathop{\rm csch}\nolimits}
\newcommand{\tridiag}{\mathop{\rm tridiag}\nolimits}
\newcommand{\AZ}{\mathop{\rm AZ}\nolimits}
\newcommand{\perm}{\mathop{\rm perm}\nolimits}
\def\hboxscript#1{ {\hbox{\scriptsize\em #1}} }
\renewcommand{\emptyset}{\varnothing}
\newcommand{\eqdef}{\stackrel{\rm def}{=}}

\newcommand{\restrict}{\upharpoonright}

\newcommand{\compinv}{{\langle -1 \rangle}}   

\newcommand{\scra}{{\mathcal{A}}}
\newcommand{\scrb}{{\mathcal{B}}}
\newcommand{\scrc}{{\mathcal{C}}}
\newcommand{\scrd}{{\mathcal{D}}}
\newcommand{\scre}{{\mathcal{E}}}
\newcommand{\scrf}{{\mathcal{F}}}
\newcommand{\scrg}{{\mathcal{G}}}
\newcommand{\scrh}{{\mathcal{H}}}
\newcommand{\scri}{{\mathcal{I}}}
\newcommand{\scrj}{{\mathcal{J}}}
\newcommand{\scrk}{{\mathcal{K}}}
\newcommand{\scrl}{{\mathcal{L}}}
\newcommand{\scrm}{{\mathcal{M}}}
\newcommand{\scrn}{{\mathcal{N}}}
\newcommand{\scro}{{\mathcal{O}}}
\newcommand{\scrp}{{\mathcal{P}}}
\newcommand{\scrq}{{\mathcal{Q}}}
\newcommand{\scrr}{{\mathcal{R}}}
\newcommand{\scrs}{{\mathcal{S}}}
\newcommand{\scrt}{{\mathcal{T}}}
\newcommand{\scrv}{{\mathcal{V}}}
\newcommand{\scrw}{{\mathcal{W}}}
\newcommand{\scrz}{{\mathcal{Z}}}

\newcommand{\bfa}{{\mathbf{a}}}
\newcommand{\bfb}{{\mathbf{b}}}
\newcommand{\bfc}{{\mathbf{c}}}
\newcommand{\bfd}{{\mathbf{d}}}
\newcommand{\bfe}{{\mathbf{e}}}
\newcommand{\bfj}{{\mathbf{j}}}
\newcommand{\bfi}{{\mathbf{i}}}
\newcommand{\bfk}{{\mathbf{k}}}
\newcommand{\bfl}{{\mathbf{l}}}
\newcommand{\bfm}{{\mathbf{m}}}
\newcommand{\bfx}{{\mathbf{x}}}
\renewcommand{\k}{{\mathbf{k}}}
\newcommand{\n}{{\mathbf{n}}}
\newcommand{\vv}{{\mathbf{v}}}
\newcommand{\bv}{{\mathbf{v}}}
\newcommand{\w}{{\mathbf{w}}}
\newcommand{\x}{{\mathbf{x}}}
\newcommand{\cc}{{\mathbf{c}}}
\newcommand{\zero}{{\mathbf{0}}}
\newcommand{\one}{{\mathbf{1}}}
\newcommand{\bmm}{{\mathbf{m}}}

\newcommand{\ahat}{{\widehat{a}}}
\newcommand{\Zhat}{{\widehat{Z}}}

\newcommand{\C}{{\mathbb C}}
\newcommand{\D}{{\mathbb D}}
\newcommand{\Z}{{\mathbb Z}}
\newcommand{\N}{{\mathbb N}}
\newcommand{\Q}{{\mathbb Q}}
\newcommand{\PP}{{\mathbb P}}
\newcommand{\R}{{\mathbb R}}
\newcommand{\RR}{{\mathbb R}}
\newcommand{\E}{{\mathbb E}}

\newcommand{\ba}{{\bm{a}}}
\newcommand{\bahat}{{\widehat{\bm{a}}}}
\newcommand{\sfa}{{{\sf a}}}
\newcommand{\bb}{{\bm{b}}}
\newcommand{\bc}{{\bm{c}}}
\newcommand{\bchat}{{\widehat{\bm{c}}}}
\newcommand{\bd}{{\bm{d}}}
\newcommand{\bee}{{\bm{e}}}
\newcommand{\bff}{{\bm{f}}}
\newcommand{\bg}{{\bm{g}}}
\newcommand{\bh}{{\bm{h}}}
\newcommand{\bll}{{\bm{\ell}}}
\newcommand{\bp}{{\bm{p}}}
\newcommand{\br}{{\bm{r}}}
\newcommand{\bs}{{\bm{s}}}
\newcommand{\bu}{{\bm{u}}}
\newcommand{\bw}{{\bm{we}}}
\newcommand{\bx}{{\bm{x}}}
\newcommand{\by}{{\bm{y}}}
\newcommand{\bz}{{\bm{z}}}
\newcommand{\bA}{{\bm{A}}}
\newcommand{\bB}{{\bm{B}}}
\newcommand{\bC}{{\bm{C}}}
\newcommand{\bE}{{\bm{E}}}
\newcommand{\bF}{{\bm{F}}}
\newcommand{\bG}{{\bm{G}}}
\newcommand{\bH}{{\bm{H}}}
\newcommand{\bI}{{\bm{I}}}
\newcommand{\bJ}{{\bm{J}}}
\newcommand{\bM}{{\bm{M}}}
\newcommand{\bN}{{\bm{N}}}
\newcommand{\bP}{{\bm{P}}}
\newcommand{\bQ}{{\bm{Q}}}
\newcommand{\bS}{{\bm{S}}}
\newcommand{\bT}{{\bm{T}}}
\newcommand{\bW}{{\bm{W}}}
\newcommand{\bX}{{\bm{X}}}
\newcommand{\bIB}{{\bm{IB}}}
\newcommand{\bOB}{{\bm{OB}}}
\newcommand{\bOS}{{\bm{OS}}}
\newcommand{\bERR}{{\bm{ERR}}}
\newcommand{\bSP}{{\bm{SP}}}
\newcommand{\bMV}{{\bm{MV}}}
\newcommand{\bBM}{{\bm{BM}}}
\newcommand{\balpha}{{\bm{\alpha}}}
\newcommand{\bbeta}{{\bm{\beta}}}
\newcommand{\bgamma}{{\bm{\gamma}}}
\newcommand{\bdelta}{{\bm{\delta}}}
\newcommand{\bkappa}{{\bm{\kappa}}}
\newcommand{\bomega}{{\bm{\omega}}}
\newcommand{\bsigma}{{\bm{\sigma}}}
\newcommand{\btau}{{\bm{\tau}}}
\newcommand{\bpsi}{{\bm{\psi}}}
\newcommand{\bzeta}{{\bm{\zeta}}}
\newcommand{\bone}{{\bm{1}}}
\newcommand{\bzero}{{\bm{0}}}

\newcommand{\Cbar}{{\overline{C}}}
\newcommand{\Dbar}{{\overline{D}}}
\newcommand{\dbar}{{\overline{d}}}
\def\Ctilde{{\widetilde{C}}}
\def\Etilde{{\widetilde{E}}}
\def\Ftilde{{\widetilde{F}}}
\def\Gtilde{{\widetilde{G}}}
\def\Htilde{{\widetilde{H}}}
\def\Ptilde{{\widetilde{P}}}
\def\Chat{{\widehat{C}}}
\def\ctilde{{\widetilde{c}}}
\def\zbar{{\overline{Z}}}
\def\pitilde{{\widetilde{\pi}}}

\newcommand{\lambdatilde}{\lambda_{\rm halt}}
\newcommand{\lambdahat}{\lambda_{\rm fly}}


\newenvironment{sarray}{
             \textfont0=\scriptfont0
             \scriptfont0=\scriptscriptfont0
             \textfont1=\scriptfont1
             \scriptfont1=\scriptscriptfont1
             \textfont2=\scriptfont2
             \scriptfont2=\scriptscriptfont2
             \textfont3=\scriptfont3
             \scriptfont3=\scriptscriptfont3
           \renewcommand{\arraystretch}{0.7}
           \begin{array}{l}}{\end{array}}

\newenvironment{scarray}{
             \textfont0=\scriptfont0
             \scriptfont0=\scriptscriptfont0
             \textfont1=\scriptfont1
             \scriptfont1=\scriptscriptfont1
             \textfont2=\scriptfont2
             \scriptfont2=\scriptscriptfont2
             \textfont3=\scriptfont3
             \scriptfont3=\scriptscriptfont3
           \renewcommand{\arraystretch}{0.7}
           \begin{array}{c}}{\end{array}}


\section{Introduction}

\begin{figure}[b]
\begin{center}
\vspace*{-4mm}
\includegraphics[width=0.45\textwidth]{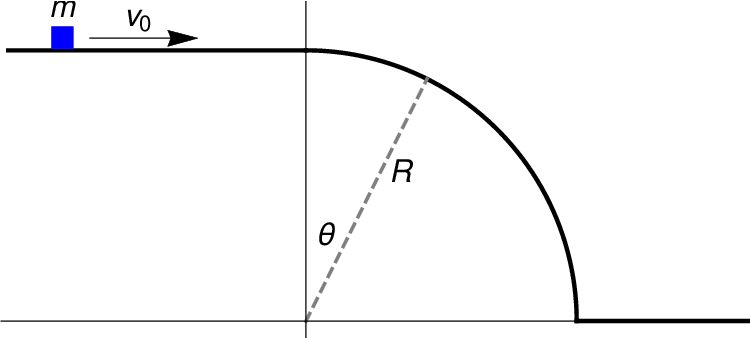}
\vspace*{-3mm}
\end{center}
   \caption{Skier on a hill of quarter-circular cross section.
            The horizontal portion of the hill is frictionless;
            the circular portion has a coefficient of kinetic friction $\mu$.
           }
   \label{fig.skier}
\end{figure}

\begin{figure}[b]
\begin{center}
\vspace*{-2mm}
\includegraphics[width=0.4\textwidth]{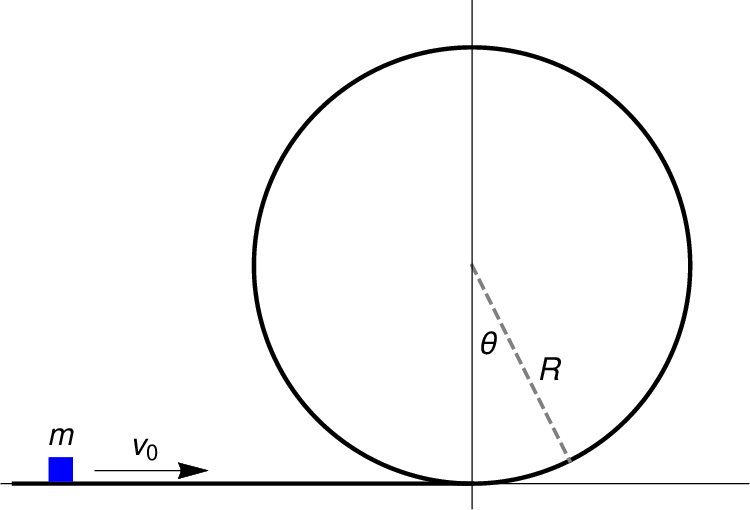}
\vspace*{-3mm}
\end{center}
   \caption{Particle on a loop-the-loop.
            The horizontal portion of the track is frictionless;
            the circular portion has a coefficient of kinetic friction $\mu$.
           }
   \label{fig.loop-the-loop}
\end{figure}

Two classic homework exercises in an elementary mechanics course
are the skier on a hemispherical hill (Fig.~\ref{fig.skier})
and the particle on a loop-the-loop track
(Fig.~\ref{fig.loop-the-loop}).\cite{mechanics_books}
Both problems illustrate nicely the use of conservation of energy
(to find the speed as a function of height)
followed by ${\bf F} = m{\bf a}$
(to find the normal force).

It is interesting to consider what happens when
the hill or track is endowed with a coefficient of kinetic friction $\mu$.
Somewhat surprisingly, the exact differential equations
turn out to be analytically solvable.\cite{Franklin_80,Mania_02,Mungan_03,%
Hite_04,Prior_07,DeLange_08,Klobus_11,Nahin_15,Gonzalez-Cataldo_17,DelPino_18}
Our purpose here is to provide a unified treatment of the two problems,
using only elementary methods that are easily accessible to undergraduates
(e.g.\ linear first-order differential equations).
Though most of our results have been obtained previously
--- as we shall document in detail ---
they are somewhat scattered in the literature.
It may thus be of some modest value to have a complete elementary derivation
collected in one place.

The skier and loop-the-loop problems
give rise to very similar differential equations,
which differ only by some sign changes.
However, these sign changes lead to significant differences
in the qualitative interpretation of the solutions.
Since the skier problem turns out to be somewhat simpler,
we treat it first and give a complete solution;
in particular, we determine the exact ``phase diagram''
in the two-dimensional parameter plane.
For the loop-the-loop, we solve the differential equations
only up to the first time (if any) that the particle halts
or completes one cycle of the loop,
so we obtain only a partial ``phase diagram''.
The full phase diagram will (as we explain later)
contain an infinite sequence of bifurcations,
and we leave its computation to a reader who wishes to take up
where we have left off.

\section{Skier on a hemispherical hill}

Consider a skier of mass $m$ on a hemispherical hill of radius $R$
(or more generally, any hill of circular cross section)
and coefficient of kinetic friction $\mu$,
entering at the top with forward velocity $v_0$;
let $\theta$ denote the angle from the vertical (Fig.~\ref{fig.skier}).
Then the radial and tangential components
of ${\bf F} = m{\bf a}$ are\cite{F=ma}
\begin{eqnarray}
   N - mg\cos\theta  & = &  -mR \dot{\theta}^2
       \label{eq.skier.radial}  \\[2mm]
   mg\sin\theta - \mu N \sgn(\dot{\theta})   & = &   mR \ddot{\theta}
       \label{eq.skier.tangential}
\end{eqnarray}
This is a pair of coupled differential equations for the unknown functions
$\theta(t)$ and $N(t)$.
We stress, however, that these equations are valid only as long as $N \ge 0$;
after that, the skier flies off the hill.
Since it is clear that the skier will only go down the hill, not up,
we have $\dot{\theta} \ge 0$ throughout the motion,
and the factor $\sgn(\dot{\theta})$ in Eq.~\reff{eq.skier.tangential}
can be dropped.\cite{note_static_friction}

Differentiating Eq.~\reff{eq.skier.radial} with respect to time yields
\be
   {dN \over dt}
   \;=\;
   - (mg\sin\theta + 2mR\ddot{\theta}) \, \dot{\theta}
   \;,
 \label{eq.skier.eq3}
\ee
and inserting $\ddot{\theta}$ from Eq.~\reff{eq.skier.tangential}
[with $\sgn(\dot{\theta}) = 1$]
yields
\be
   {dN \over dt}
   \;=\;
   - (3mg\sin\theta - 2\mu N) \, \dot{\theta}
   \;.
 \label{eq.skier.eq4}
\ee
Using the chain rule $dN/dt = (dN/d\theta) (d\theta/dt)$
we can eliminate $\dot{\theta}$ from Eq.~\reff{eq.skier.eq4},
leading to
\be
   {dN \over d\theta}  \,-\, 2\mu N  \;=\;  -3mg\sin\theta
   \;.
 \label{eq.diffeqn.Ntheta}
\ee
This is a first-order inhomogeneous linear differential equation
with constant coefficients for the unknown function $N(\theta)$,
and it can be solved by the method of integrating factors.
Here the integrating factor is $e^{-2\mu\theta}$,
and the solution is\cite{alternate_solution}
\be
   N(\theta)
   \;=\;
   N_0 e^{2\mu\theta}
   \:-\: 3mg \,
   {e^{2\mu\theta} - \cos\theta - 2\mu\sin\theta  \over  1 + 4\mu^2}
 \label{eq.soln.Ntheta.0}
\ee
where $N_0 = N(0)$.
We again stress that this solution is valid only where $N(\theta) \ge 0$;
at the first angle (if~any) where $N(\theta)$ crosses zero
to a negative value, the skier flies off the hill.

Evaluating Eq.~\reff{eq.skier.radial} at $\theta = 0$,
where the skier's angular velocity is $\dot{\theta} = v_0/R$,
we obtain $N_0 = mg - mv_0^2/R$.
In particular, if the dimensionless parameter $\lambda \eqdef v_0^2/gR$
is $\ge 1$, then $N_0 \le 0$ and the skier immediately flies off the hill;
we therefore assume henceforth that $0 \le \lambda < 1$.
Inserting $N_0 = (1-\lambda)mg$ in Eq.~\reff{eq.soln.Ntheta.0}, we obtain
\be
   N(\theta)
   \;=\;
   (1-\lambda)mg \, e^{2\mu\theta}
   \:-\: 3mg \,
   {e^{2\mu\theta} - \cos\theta - 2\mu\sin\theta  \over  1 + 4\mu^2}
   \;,
 \label{eq.soln.Ntheta}
\ee
which is the closed-form solution
giving the normal force as a function of angle.

In the absence of friction ($\mu = 0$), Eq.~\reff{eq.soln.Ntheta} simplifies to
\be
   N(\theta)  \;=\;  (3\cos\theta - 2 - \lambda) mg
   \;.
\ee
This is a decreasing function of $\theta$,
and skier flies off the hill when $N = 0$, i.e.\ when
\be
   \theta  \;=\;  \cos^{-1} \Big( {2+\lambda \over 3} \Big)
   \;.
 \label{eq.thetafly.mu=0}
\ee
In the usual textbook problem one has also $v_0 = 0$ (i.e.\ $\lambda = 0$),
and we obtain the standard answer that the skier flies off at angle
$\theta = \cos^{-1}(2/3) \approx 48.19^\circ$.
   
When $\mu > 0$, by contrast, the normal force is no longer
a decreasing function of $\theta$,
nor is it guaranteed to reach zero within the interval
$0 \le \theta \le \pi/2$.
Indeed, $dN/d\theta |_{\theta = 0} = 2\mu (1-\lambda)mg > 0$,
so the normal force is initially increasing.

We can also obtain the velocity as a function of angle.
It is convenient to define the dimensionless quantity
$\Lambda \eqdef v^2/gR = R \dot{\theta}^2/g$;
its value at $\theta=0$ is what we have called $\lambda$.
Then from Eq.~\reff{eq.skier.radial} we have immediately
\be
   N  \;=\;  (\cos\theta - \Lambda) mg
 \label{eq.N.Lambda}
\ee
[which reduces to $N_0 = (1-\lambda)mg$ when $\theta=0$] or equivalently
\be
   \Lambda   \;=\;  \cos \theta \,-\, {N \over mg} \;.
 \label{eq.Lambda.N}
\ee
In particular, from $N \ge 0$ we deduce that $\Lambda \le \cos\theta$:
 this gives the maximum speed that the skier can have at any given angle
 if she is to avoid flying off the hill.
Combining Eqs.~\reff{eq.soln.Ntheta} and \reff{eq.Lambda.N} gives
the closed-form solution for the
speed as a function of angle:\cite{speed_as_a_function_of_angle}
\be
   \Lambda(\theta)
   \;=\;
   \cos\theta  \:-\:
   (1-\lambda) \, e^{2\mu\theta}
   \:+\: 3 \,
   {e^{2\mu\theta} - \cos\theta - 2\mu\sin\theta  \over  1 + 4\mu^2}
   \;.
 \label{eq.soln.Lambdatheta}
\ee
Note, however, that this solution is valid only where $\Lambda(\theta) \ge 0$;
at the first angle (if~any) where $\Lambda(\theta) = 0$,
the skier comes to rest (perhaps only asymptotically as $t \to +\infty$).
The solution \reff{eq.soln.Lambdatheta}
must therefore be supplemented by the two inequalities
$0 \le \Lambda(\theta) \le \cos\theta$.

{}From Eqs.~\reff{eq.diffeqn.Ntheta} and \reff{eq.N.Lambda}/\reff{eq.Lambda.N}
we see that $\Lambda(\theta)$
satisfies the differential equation\cite{differential_equation}
\be
   {d\Lambda \over d\theta}  \,-\, 2\mu \Lambda
   \;=\;
   2 (\sin\theta \,-\, \mu \cos\theta)
   \;.
 \label{eq.diffeqn.Lambdatheta}
\ee
The solution of this differential equation
with the initial condition $\Lambda(0) = \lambda$
is of course Eq.~\reff{eq.soln.Lambdatheta}.\cite{alternate_solution}

In the absence of friction ($\mu = 0$), Eq.~\reff{eq.soln.Lambdatheta}
simplifies to
\be
   \Lambda(\theta)  \;=\;  \lambda \:+\: 2(1 - \cos\theta)
   \;,
\ee
which is just the expression for conservation of energy:
$\half mv^2 = \half m v_0^2 + mgR(1-\cos\theta)$.
More generally, the kinetic energy plus gravitational potential energy is
\be
   E  \;=\;  \half mv^2 \,+\, mgR(\cos\theta - 1)
      \;=\;  \half mgR \bigl[ \Lambda(\theta) \,+\, 2(\cos\theta - 1) \bigr]
   \;,
 \label{eq.energy}
\ee
so that
\be
   {dE \over dt}
   \;=\;
   \half mgR \Bigl[ {d\Lambda \over d\theta} \,-\, 2\sin\theta \Bigr]
       \dot{\theta}
   \;.
 \label{eq.dEdt}
\ee
The work-energy theorem asserts that $dE/dt$ must equal the
rate of work done by friction, which is $-\mu N R \dot{\theta}$;
and this equality is an immediate consequence of
Eqs.~\reff{eq.N.Lambda}, \reff{eq.diffeqn.Lambdatheta}
and \reff{eq.dEdt}.
Conversely, the differential equation \reff{eq.diffeqn.Lambdatheta}
 could alternatively be derived by combining the work-energy theorem
 with Eqs.~\reff{eq.N.Lambda} and \reff{eq.dEdt}.\cite{work-energy}
It may be useful for students to compare these two derivations:
one directly from the Newtonian equations of motion,
the other from the work-energy theorem.

Finally, we can use Eq.~\reff{eq.soln.Lambdatheta} to obtain the
time-dependence of the motion.
{}From $\Lambda = R \dot{\theta}^2/g$ we have
\be
   {d\theta \over dt}
   \;=\;
   \Bigl[ {g \over R} \, \Lambda(\theta) \Bigr]^{1/2}
\ee
and hence
\be
   t(\theta)
   \;=\;
   \int\limits_0^\theta
      {d\theta' \over \Bigl[ \displaystyle {g \over R} \,
                             \Lambda(\theta') \Bigr]^{1/2}}
   \;.
 \label{eq.time}
\ee

%
%

\begin{figure}[p]
\begin{center}
\includegraphics[width=0.8\textwidth]{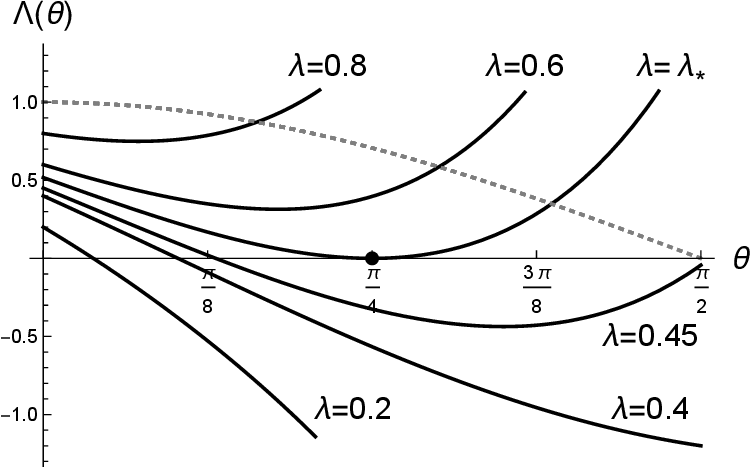}
\end{center}
   \caption{The curves $\Lambda(\theta)$ for $\mu = 1$ and
            $\lambda = 0.2$, 0.4, 0.45, $\lambda_\star(1) \approx 0.517594$,
            0.6 and 0.8.
            The skier halts when $\Lambda(\theta) = 0$,
            or flies off the hill when $\Lambda(\theta) = \cos\theta$
            (shown as a dotted curve),
            whichever happens first.
            The critical curve is $\lambda = \lambda_\star$.
            The dot indicates the point $\theta = \theta_\star$
            (here $\theta_\star = \arctan 1 = \pi/4$).
           }
   \label{fig.Lambda.theta}
\end{figure}

\begin{figure}[p]
\begin{center}
\includegraphics[width=0.8\textwidth]{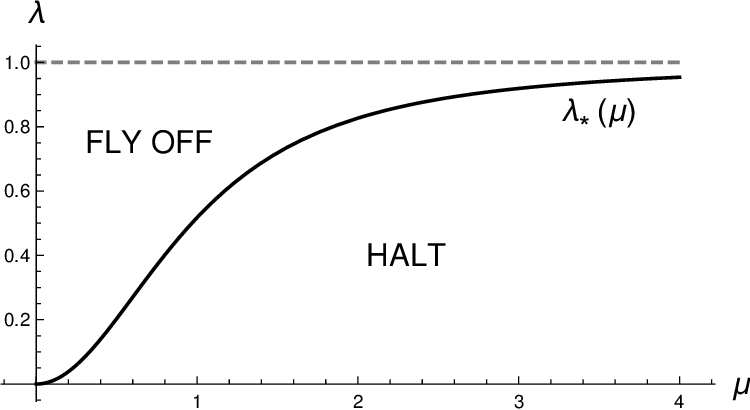}
\end{center}
   \caption{The curve $\lambda_\star(\mu)$ that forms the boundary between
               the ``halt'' phase and the ``fly-off'' phase.
           }
   \label{fig.lambdastar}
\end{figure}

We can now analyze the qualitative behavior of the motion
as a function of the two parameters
$\mu \in [0,\infty)$ and $\lambda \in [0,1)$.
We have seen that the skier halts when $\Lambda(\theta) = 0$,
or flies off the hill when $\Lambda(\theta) = \cos\theta$,
whichever happens first;
if neither happens for $\theta < \pi/2$,
then the skier reaches the bottom of the hill.
(We will see later that this last case never occurs.)
The critical solution that separates these two scenarios
is given by the trajectory for which the skier halts
at an angle $\theta_\star$ (hence $\Lambda(\theta_\star) = 0$)
that also satisfies $\Lambda'(\theta_\star) = 0$:
see the curve marked $\lambda = \lambda_\star$ in Fig.~\ref{fig.Lambda.theta}.
Applying this condition in Eq.~\reff{eq.diffeqn.Lambdatheta}
leads immediately to\cite{ref_thetastar}
\be
   \theta_\star(\mu)  \;\eqdef\;  \arctan\mu
   \;.
\ee
Substituting this in Eq.~\reff{eq.soln.Lambdatheta},
we obtain the relationship between the initial velocity
and the friction coefficient
that defines the phase boundary:\cite{lambdastar_mu}
\be
   \lambda_\star(\mu)
   \;\eqdef\;
   {4\mu^2 - 2 + 2 e^{-2\mu \arctan \mu} \sqrt{1 + \mu^2}
    \over
    1 + 4\mu^2
   }
   \;.
 \label{def.lambdastar}
\ee
Please observe that $\lambda_\star(\mu)$ is an increasing function of $\mu$
that runs from 0 to 1 as $\mu$ runs from 0 to $\infty$
(see Fig.~\ref{fig.lambdastar}).

In this way we have obtained a ``phase diagram''
that divides the $(\mu,\lambda)$ plane
into three possible qualitative behaviors:
\begin{itemize}
   \item  For $0 \le \lambda < \lambda_\star(\mu)$,
      the skier halts after a finite time
      at some angle $\theta_{\rm halt}(\mu,\lambda)$:
      this angle is an increasing function of $\lambda$
      that runs from 0 to $\arctan\mu$
      as $\lambda$ runs from 0 to $\lambda_\star(\mu)$.
   \item  For $\lambda = \lambda_\star(\mu)$,
      the skier comes to rest asymptotically as $t \to +\infty$
      at the angle $\theta=\arctan\mu$.\cite{note_equilibrium}
   \item  For $\lambda_\star(\mu) < \lambda < 1$,
      the skier flies off the hill at some angle
      $\theta_{\rm fly}(\mu,\lambda)$:
      this angle is a decreasing function of $\lambda$
      that tends to 0 as $\lambda \to 1$.
\end{itemize}
\nopagebreak
The curve $\lambda_\star(\mu)$ thus forms the boundary between
the ``halt'' phase and the ``fly-off'' phase
(see again Fig.~\ref{fig.lambdastar}).\cite{phase_diagram}
In particular, the skier always either halts or flies off;
she never reaches angle $\pi/2$.

\begin{figure}[p]
\begin{center}
\includegraphics[width=0.8\textwidth]{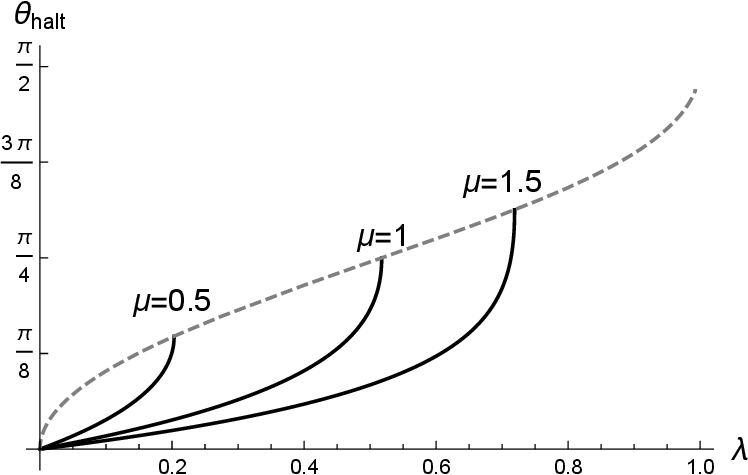}
\end{center}
   \caption{$\theta_{\rm halt}$ as a function of $\lambda$
               in the ``halt'' phase $0 \le \lambda \le \lambda_\star(\mu)$,
            for $\mu = 0.5$, 1, 1.5.
            The endpoints lie on the dashed curve, defined parametrically by
               $\lambda = \lambda_\star(\mu)$ and $\theta = \arctan \mu$.
           }
   \label{fig.thetahalt}
\end{figure}

\begin{figure}[p]
\begin{center}
\includegraphics[width=0.8\textwidth]{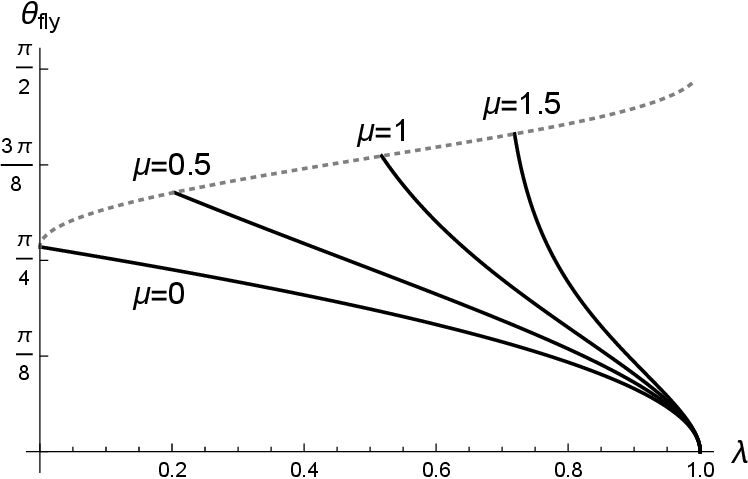}
\end{center}
   \caption{$\theta_{\rm fly}$ as a function of $\lambda$
               in the ``fly-off'' phase $\lambda_\star(\mu) < \lambda \le 1$,
            for $\mu = 0$, 0.5, 1, 1.5.
            The endpoints lie on the dotted curve,
            corresponding to $\lambda \to \lambda_\star(\mu)$ from above.
            For~$\mu = 0$ we have the closed-form solution
            \reff{eq.thetafly.mu=0}.
           }
   \label{fig.thetafly}
\end{figure}

Some typical curves of $\Lambda(\theta)$ for all three scenarios
are shown in Fig.~\ref{fig.Lambda.theta}.
Note in particular that $\Lambda(\theta) = \Lambda'(\theta) = 0$
when $\lambda = \lambda_\star(\mu)$ and $\theta = \theta_\star(\mu)$;
and note the fundamental qualitative difference
between the curves for $\lambda < \lambda_\star(\mu)$,
which reach the $\Lambda = 0$ axis,
and those for $\lambda > \lambda_\star(\mu)$,
which do not.\cite{refs_for_fig.Lambda.theta}

Some typical curves of $\theta_{\rm halt}(\mu,\lambda)$
as a function of $\lambda$ are shown in Fig.~\ref{fig.thetahalt},
and some typical curves of $\theta_{\rm fly}(\mu,\lambda)$
as a function of $\lambda$ are shown
in Fig.~\ref{fig.thetafly}.
Please note the discontinuous change in behavior
as the phase boundary $\lambda_\star(\mu)$ is crossed:
$\theta_{\rm fly}(\mu,\lambda_\star(\mu))$
[the dotted curve in Fig.~\ref{fig.thetafly}]
is much larger than $\theta_{\rm halt}(\mu,\lambda_\star(\mu))$
[the dashed curve in Fig.~\ref{fig.thetahalt}].\cite{ref_for_figs_thetahalt+thetafly}
This is a very simple example of sensitive dependence
to initial conditions, giving rise to a discontinuous phase transition ---
a phenomenon pointed out already
by James Clerk Maxwell in 1876.\cite{ref_Maxwell}

Since the proofs of all the previous claims
involve some slightly intricate calculus,
we relegate them to Appendix~A
in the Supplementary Materials.\cite{ref_supplementary}

Let us remark, finally, that by the same methods one can study
the more general problem in which the coefficient of kinetic friction
is an arbitrary function $\mu(\theta)$ of the position along the hill:
the equation \reff{eq.diffeqn.Ntheta}
is still a first-order inhomogeneous linear differential equation
for the unknown function $N(\theta)$ ---
albeit now one with nonconstant coefficients ---
so can still be solved by the method of integrating factors
(though the result may not be analytically expressible
in terms of elementary functions).
We leave it to interested readers to pursue this generalization.

Some recent related articles are
Refs.~\onlinecite{Gonzalez-Cataldo_17}, \onlinecite{DelPino_18}
and \onlinecite{Ivchenko_21},
which study a particle sliding down an arbitrary curve in the presence
of kinetic friction;
Ref.~\onlinecite{Balart_19},
which uses the Lagrangian formalism with Lagrange multipliers
to analyze a particle sliding {\em without friction}\/
down an arbitrary concave curve;
and Ref.~\onlinecite{Mejia_20},
which studies a ball rolling
(initially without slipping, later with sliding and kinetic friction)
on an arbitrary curve in the presence of gravity,
including an experimental realization.


\section{Particle on loop-the-loop track}

A block of mass $m$ is injected with forward velocity $v_0$
into a loop-the-loop track of radius $R$
and coefficient of kinetic friction $\mu$;
let $\theta$ denote the angle up from the bottom,
as shown in Fig.~\ref{fig.loop-the-loop}.
(In one common version of the problem\cite{mechanics_books},
 the block is released from rest at height $h$
 and slides to the bottom via a frictionless track;
 in this case $v_0 = \sqrt{2gh}$.)
The radial and tangential components of ${\bf F} = m{\bf a}$ are
\begin{eqnarray}
   mg\cos\theta - N  & = &  -mR \dot{\theta}^2
       \label{eq.loop.radial}  \\[2mm]
   -mg\sin\theta - \mu N \sgn(\dot{\theta})   & = &   mR \ddot{\theta}
       \label{eq.loop.tangential}
\end{eqnarray}
As before, these equations are valid only as long as $N \ge 0$;
after that, the block falls off the track.

The loop-the-loop problem is more complicated than the skier,
for three reasons:
the particle can cycle around the track;
it can reverse direction;
and it can halt due to static friction.
Each time the particle reverses direction,
we need to apply Eq.~\reff{eq.loop.tangential}
with a new value for $\sgn(\dot{\theta})$;
this repeated switching between different equations seems quite complicated,
and probably needs to be handled by numerical solution.\cite{loop-the-loop_reverses}
To simplify matters, we will here follow the block
only until it first reaches $\dot{\theta} = 0$ or falls off the track;
we therefore have $\dot{\theta} \ge 0$.

Proceeding as in Eqs.~\reff{eq.skier.eq3}--\reff{eq.diffeqn.Ntheta}
leads to the differential equation
\be
   {dN \over d\theta}  \,+\, 2\mu N  \;=\;  -3mg\sin\theta
 \label{eq.diffeqn.Ntheta.loop}
\ee
for the unknown function $N(\theta)$;
this equation differs from Eq.~\reff{eq.diffeqn.Ntheta}
only by the replacement $\mu \to -\mu$.
The solution is therefore
\be
   N(\theta)
   \;=\;
   N_0 e^{-2\mu\theta}
   \:-\: 3mg \,
   {e^{-2\mu\theta} - \cos\theta + 2\mu\sin\theta  \over  1 + 4\mu^2}
 \label{eq.soln.Ntheta.0.loop}
\ee
where $N_0 = N(0)$.
Applying Eq.~\reff{eq.loop.radial} at $\theta = 0$,
where the block's angular velocity is $\dot{\theta} = v_0/R$,
we see that $N_0 = mg + mv_0^2/R$.
Using again the dimensionless parameter $\lambda \eqdef v_0^2/gR$,
we have $N_0 = (1+\lambda)mg$ and hence
\be
   N(\theta)
   \;=\;
   (1+\lambda)mg \, e^{-2\mu\theta}
   \:-\: 3mg \,
   {e^{-2\mu\theta} - \cos\theta + 2\mu\sin\theta  \over  1 + 4\mu^2}
   \;.
 \label{eq.soln.Ntheta.loop}
\ee

To obtain the velocity as a function of angle,
we define once again the dimensionless quantity
$\Lambda \eqdef v^2/gR = R \dot{\theta}^2/g$,
which takes the value $\lambda$ at $\theta=0$.
Then from Eq.~\reff{eq.loop.radial} we have
\be
   N  \;=\;  (\cos\theta + \Lambda) mg
\ee
[which reduces to $N_0 = (1+\lambda)mg$ when $\theta=0$]
and therefore\cite{loop-the-loop_Lambda_theta}
\be
   \Lambda(\theta)
   \;=\;
   -\cos\theta
   \:+\:
   (1+\lambda) \, e^{-2\mu\theta}
   \:-\: 3 \,
   {e^{-2\mu\theta} - \cos\theta + 2\mu\sin\theta  \over  1 + 4\mu^2}
   \;.
 \label{eq.soln.Lambdatheta.loop}
\ee
Since $\Lambda \ge 0$, we must have $N \ge mg\cos\theta$;
and when $N = mg\cos\theta$, the block comes instantaneously to rest.
After that, the particle might either reverse direction
or halt due to static friction.
As mentioned earlier, we refrain from following the particle
beyond the first time it comes instantaneously to rest.

The solution \reff{eq.soln.Ntheta.loop}
must therefore be supplemented by the two inequalities
$N(\theta) \ge 0$ and $N(\theta) \ge mg\cos\theta$.
(Please note that, unlike in the skier problem,
 both of these inequalities point in the {\em same}\/ direction;
 this radically changes the nature of the qualitative analysis.)
The block comes instantaneously to rest when $N(\theta) = mg\cos\theta$,
or falls off the track when $N(\theta) = 0$, whichever happens first;
if neither happens for $\theta < 2\pi$,
then the block completes one full cycle of the loop-the-loop.
Now, the inequality $N(\theta) \ge mg\cos\theta$
is the more stringent one in the lower half of the loop-the-loop
(that is, $-\pi/2 \le \theta \le \pi/2$ modulo $2\pi$),
while the inequality $N(\theta) \ge 0$
is the more stringent one in the upper half of the loop-the-loop
(that is, $\pi/2 \le \theta \le 3\pi/2$ modulo $2\pi$).
Therefore, the block can come instantaneously to rest
only in the lower half of the loop-the-loop,
and it can fall off the track only in the upper half of the loop-the-loop.

In the absence of friction ($\mu = 0$), Eq.~\reff{eq.soln.Ntheta.loop}
simplifies to
\be
   N(\theta)  \;=\;  (\lambda - 2 + 3\cos\theta) mg
   \;.
\ee
If $\lambda \le 2$, then the block reverses direction at
\be
   \theta  \;=\; \theta_{\rm max} \;\eqdef\;
      \cos^{-1} \Big( {2-\lambda \over 2} \Big)
\ee
(a value that follows immediately from conservation of energy)
and oscillates forever between $-\theta_{\rm max}$ and $\theta_{\rm max}$;
if $2 < \lambda < 5$, then the block falls off the track at
\be
   \theta  \;=\; \theta_{\rm fall} \;\eqdef\;
      \cos^{-1} \Big( {2-\lambda \over 3} \Big)
   \;,
\ee
which lies between $\pi/2$ and $\pi$;
if $\lambda = 5$, then the block asymptotically approaches
$\theta = \pi$ as $t \to +\infty$;
if $\lambda > 5$,
then the block cycles forever around the track without loss of energy.

In the presence of friction ($\mu > 0$),
the analysis proceeds as follows:

\medskip

1) The first step is to determine the conditions under which
the particle halts in the first quadrant ($0 \le \theta \le \pi/2$).
The particle halts at angle $\theta$ when $\Lambda(\theta) = 0$,
i.e.\ in~case the initial velocity satisfies
\be
   \lambda
   \;=\;
   \lambdatilde(\theta,\mu)
   \;\eqdef\;
   {2 \over 1 + 4\mu^2}
   \: 
   \Bigl[ (1 - 2\mu^2) \,+\, e^{2\mu\theta}
           \bigl[ 3 \mu \sin\theta \,-\, (1-2\mu^2) \cos\theta \bigr]
   \Bigr]
   \;.
 \label{def.lambdatilde}
\ee
Since
\be
   {\partial \lambdatilde(\theta,\mu) \over \partial \theta}
   \;=\;
   2 e^{2\mu\theta} (\mu\cos\theta + \sin\theta)
   \;,
 \label{eq.lambdatilde.deriv}
\ee
$\lambdatilde(\theta,\mu)$ is an increasing function of $\theta$
in the interval $0 \le \theta \le \pi/2$
(as is intuitively clear:
 to reach a larger angle, more initial velocity is needed).
In particular, the particle reaches $\theta = \pi/2$
with $\dot{\theta} > 0$ if and only if
\be
   \lambda
   \;>\;
   \lambdatilde(\pi/2,\mu)
   \;\eqdef\;
   {2 - 4\mu^2 + 6\mu e^{\pi\mu} \over 1 + 4\mu^2}
   \;.
\ee

\begin{figure}[t]
\begin{center}
\begin{tabular}{c@{\qquad}c}
   \includegraphics[width=0.48\textwidth]{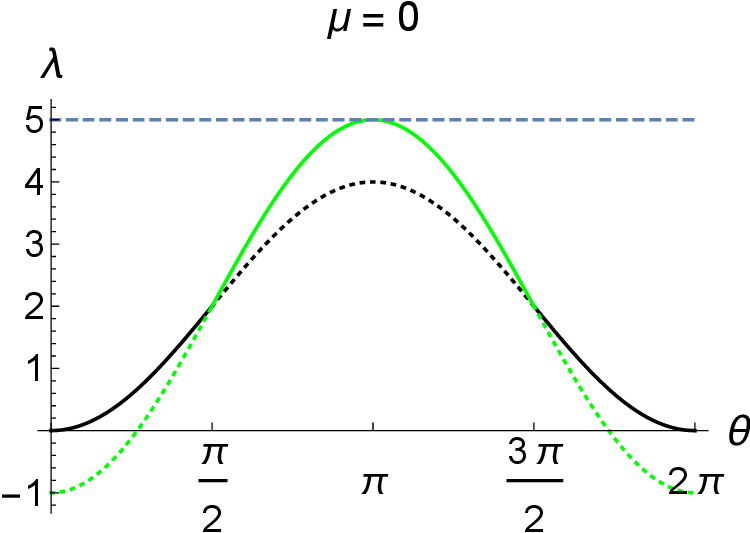} &
   \includegraphics[width=0.48\textwidth]{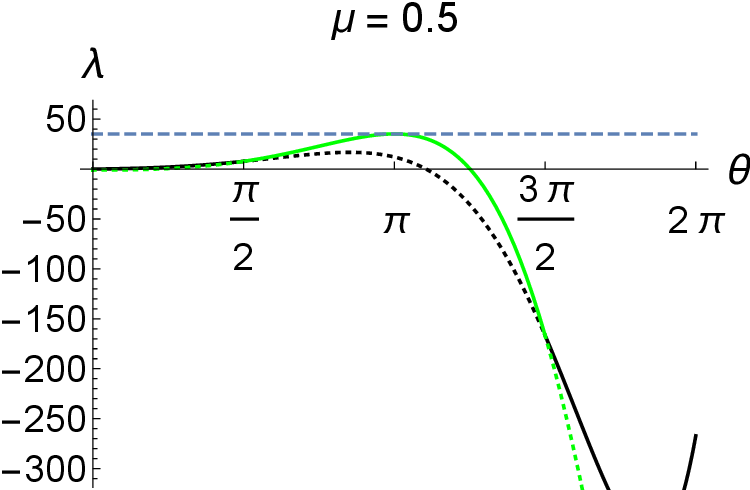} \\[1cm]
   \includegraphics[width=0.48\textwidth]{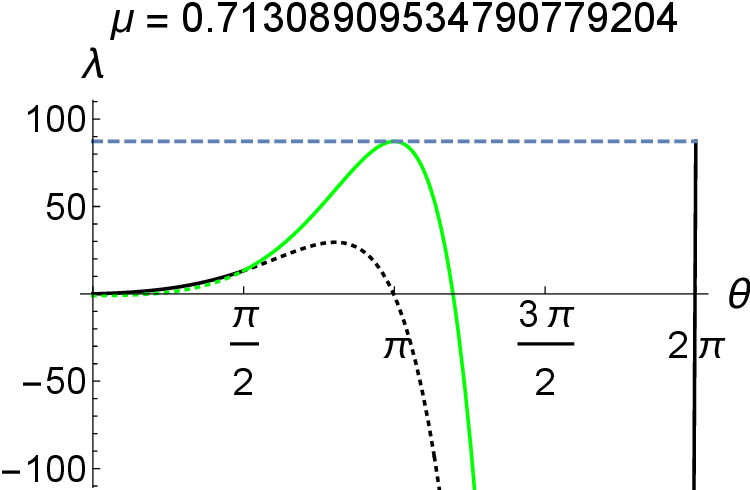} &
   \includegraphics[width=0.48\textwidth]{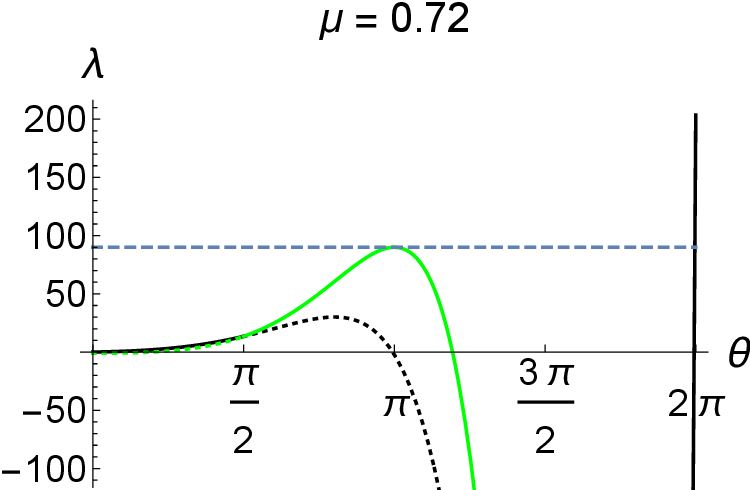}
\end{tabular}
\end{center}
   \caption{The functions $\lambdatilde$ (black)
            and $\lambdahat$ (green or gray)
            versus $\theta$ for some selected values of $\mu$.
            The dominant (resp.\ subdominant) condition
            is shown as a solid (resp.\ dotted) curve.
            A horizontal dashed line is shown at $\lambdahat(\pi,\mu)$.
            The curve in the bottom-left panel corresponds to the value
            $\mu = \mu_{\rm crit} \approx 0.713089$
            where $\lambdahat(\pi,\mu) = \lambdatilde(2\pi,\mu)$.
            \hfill\break\hspace*{4mm}
            From Eq.~\reff{eq.lambdatilde.deriv} we see that
            $\lambdatilde$ is
            increasing for $0 \le \theta \le \pi - \arctan\mu$,
            decreasing for $\pi - \arctan\mu \le \theta \le 2\pi - \arctan\mu$,
            and increasing for $2\pi - \arctan\mu \le \theta \le 2\pi$.
            From Eq.~\reff{eq.lambdahat.deriv} we see that
            $\lambdahat$ is
            increasing for $0 \le \theta \le \pi$
            and  decreasing for $\pi \le \theta \le 2\pi$.
            The two curves cross at $\pi/2$ and $3\pi/2$.
           }
   \label{fig.lambdatildehatplots}
\end{figure}

2) If the particle reaches angle $\pi/2$ without halting,
the next step is to determine the conditions under which
the particle flies off in the second or third quadrant
($\pi/2 \le \theta \le 3\pi/2$).
The particle flies off at angle $\theta$ when $N(\theta) = 0$,
i.e.\ in~case the initial velocity satisfies 
\be
   \lambda
   \;=\;
   \lambdahat(\theta,\mu)
   \;\eqdef\;
   {2 - 4 \mu^2 + 3 e^{2\mu\theta} (2\mu\sin\theta - \cos\theta)
   \over
    1 + 4\mu^2
   }
   \;.
 \label{def.lambdahat}
\ee
Note that $\lambdahat(\pi/2,\mu) = \lambdatilde(\pi/2,\mu)$.
Since
\be
   {\partial \lambdahat(\theta,\mu) \over \partial \theta}
   \;=\;
   3 e^{2\mu\theta} \sin\theta
   \;,
 \label{eq.lambdahat.deriv}
\ee
we see that $\lambdatilde(\theta,\mu)$ is an increasing function of $\theta$
in the interval from $\pi/2$ to $\pi$,
and then a decreasing function in the interval from $\pi$ to $3\pi/2$.
The first of these facts is again intuitively clear:
to survive to a larger angle without flying off,
more initial velocity is needed.
The second fact implies that if the particle reaches angle $\pi$
without flying off
--- that is, if
\be
   \lambda
   \;\ge\;
   \lambdahat(\pi,\mu)
   \;\eqdef\;
   {2 - 4 \mu^2 + 3 e^{2\pi\mu}
   \over
    1 + 4\mu^2
   }
\ee
--- then it also reaches angle $3\pi/2$ without flying off.
This is intuitively clear when there is no friction,
but not so obvious in the presence of friction.
This implies --- analogously to what happens in the skier problem ---
a discontinuous change of behavior
as $\lambda$ passes through $\lambdahat(\pi,\mu)$.
See Fig.~\ref{fig.lambdatildehatplots}
for plots of $\lambdatilde(\theta,\mu)$ and $\lambdahat(\theta,\mu)$
versus $\theta$ for some selected values of $\mu$.

3) If the particle reaches angle $\pi$ (and hence also angle $3\pi/2$)
without halting or flying off,
the next step is to determine what happens in the fourth quadrant
($3\pi/2 < \theta < 2\pi$).
The particle halts at angle $\theta$ in~case
$\lambda$ equals the quantity $\lambdatilde(\theta,\mu)$
defined in Eq.~\reff{def.lambdatilde}.
From Eq.~\reff{eq.lambdatilde.deriv}
we see that $\partial \lambdatilde(\theta,\mu)/\partial \theta$
is negative at $\theta = 3\pi/2$ and positive at $\theta = 2\pi$,
with a unique zero at $\theta = 2\pi - \arctan\mu$.
So $\lambdatilde(\theta,\mu)$
is decreasing in the interval $3\pi/2 \le \theta \le 2\pi - \arctan\mu$
and increasing in the interval $ 2\pi - \arctan\mu \le \theta \le 2\pi$.
Its maximum value in the interval $[3\pi/2,2\pi]$
therefore lies either at $\theta = 3\pi/2$ or at $\theta = 2\pi$.
Since we are in the situation
$\lambda \ge \lambdahat(\pi,\mu) > \lambdahat(3\pi/2,\mu) =
                                   \lambdatilde(3\pi/2,\mu)$,
the only relevant question is whether
$\lambda$ is larger than $\lambdatilde(2\pi,\mu)$ or not.
If it is, then the particle reaches angle $2\pi$ without halting.
If it is not, then the particle halts at some angle
in the interval $(2\pi - \arctan\mu,2\pi]$,
namely, the unique angle where $\lambda = \lambdatilde(\theta,\mu)$.
The first of these cases always occurs when
$\lambdahat(\pi,\mu) > \lambdatilde(2\pi,\mu)$,
i.e.\ when $0 \le \mu < \mu_{\rm crit} \approx 0.713089$.
(See Appendix~B in the Supplementary Materials\cite{ref_supplementary}
 for the proof that there is a unique such value $\mu_{\rm crit}$.)
When $\mu \ge \mu_{\rm crit}$,
then there is a ``halt in fourth quadrant'' phase at
$\lambdahat(\pi,\mu) \le \lambda \le \lambdatilde(2\pi,\mu)$
and a ``survive to angle $2\pi$'' phase at
$\lambda \ge \lambdatilde(2\pi,\mu)$.
We record the formula
\be
   \lambdatilde(2\pi,\mu)
   \;\eqdef\;
   {(4\mu^2 - 2) (e^{4\pi\mu} - 1)  \over 1 + 4\mu^2}
   \;.
 \label{def.lambdatilde.2pi}
\ee

4) If the particle survives to angle $2\pi$,
then it has there a forward velocity corresponding to a value
\begin{subeqnarray}
   \lambda_{\rm new}
   \;\eqdef\;
   \Lambda(2\pi)
   & = &
   \lambda e^{-4\pi\mu}
   \:+\:
   {(2 - 4\mu^2) (1 - e^{-4\pi\mu}) \over  1 + 4\mu^2}
        \\[2mm]
   & = &
   e^{-4\pi\mu} \bigl[ \lambda \,-\, \lambdatilde(2\pi,\mu) \bigr]
        \\[2mm]
   & \ge &
   0   \;.
 \label{def.lambda_new}
\end{subeqnarray}
Since $\lambdatilde(2\pi,\mu) > 0$ in the ``survive to angle $2\pi$'' phase,
we have $\lambda_{\rm new} < e^{-4\pi\mu} \lambda$:
thus the kinetic energy is reduced by at least a factor $e^{-4\pi\mu}$
at each revolution.
The subsequent motion can then be found by repeating the foregoing analysis
with $\lambda$ replaced by $\lambda_{\rm new}$.

\begin{figure}[t]
\begin{center}
\includegraphics[width=0.8\textwidth]{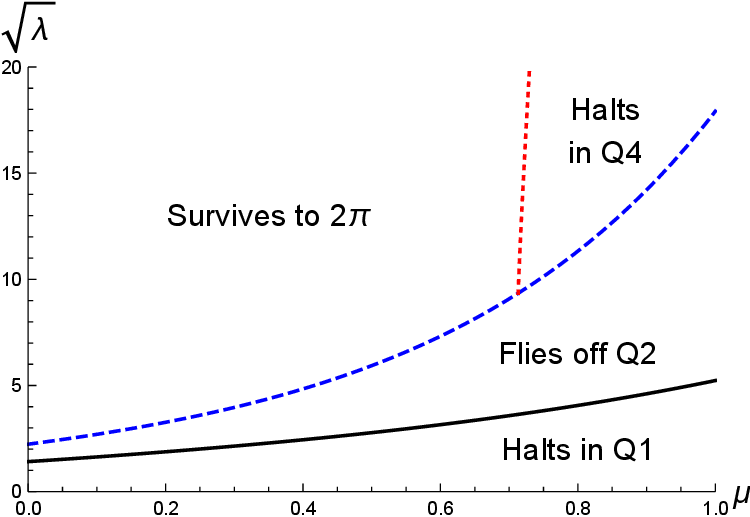}
\end{center}
   \caption{Phase diagram for the loop-the-loop problem,
            up to the first time that the particle reaches
            $\dot{\theta} = 0$ or $\theta =2\pi$.
            The vertical axis shows $\sqrt{\lambda}$.
            The three boundary curves are, from bottom to top,
            $\lambdatilde(\pi/2,\mu)$,
            $\lambdahat(\pi,\mu)$ and
            $\lambdatilde(2\pi,\mu)$,
            shown respectively in solid black, dashed blue, dotted red
            (color online).
            The particle either halts in the first quadrant (Q1),
            flies off the second quadrant (Q2),
            halts in the fourth quadrant (Q4),
            or survives to angle $2\pi$.
           }
   \label{fig.loop-the-loop_phasediag}
\end{figure}

The resulting phase diagram is shown in
Fig.~\ref{fig.loop-the-loop_phasediag}.
Since $\lambdatilde(2\pi,\mu)$ grows extremely rapidly with $\mu$,
we have used $\sqrt{\lambda}$ instead of $\lambda$ on the vertical axis,
to compress the plot.
This phase diagram agrees with the one
found by K\l{}obus (Ref.~\onlinecite{Klobus_11}, Fig.~2);
the value of $\lambdatilde(2\pi,1)$ also agrees with his.
All three phase boundaries are increasing functions of $\mu$:
see Appendices~B1--B3 in the Supplementary Materials.\cite{ref_supplementary}

Of course, this phase diagram only follows the particle
up to the first time that it reaches $\dot{\theta} = 0$ or $\theta =2\pi$.
A more complete analysis would show that
the phase ``survives to angle $2\pi$'' is itself divided into sub-phases
``halts in the first quadrant'' ($2 \pi < \theta < 5\pi/2$),
``flies off the second quadrant'' ($5\pi/2 < \theta < 3\pi$),
``halts in the fourth quadrant'' ($7\pi/2 < \theta < 4\pi$)
and ``survives to angle $4\pi$'';
and this latter phase is further divided into sub-phases;
and so on infinitely.
We leave it to interested readers to work out the details
of this infinite sequence of bifurcations.

\section*{Acknowledgments}

We are extremely grateful to three referees
for their detailed and helpful comments on several versions of this paper.

\clearpage

\begin{center}
   \Large
   {\bf Supplementary Materials for} \\[5mm]
   Skier and loop-the-loop with friction \\
   {\it Dominik Kufel and Alan D.~Sokal} \\[1cm]
\end{center}

\appendix
\renewcommand\thefigure{\thesection\arabic{figure}}
\section{Proofs for the skier}
\setcounter{figure}{0}

\medskip

\subsection{Behavior of the function $\bm{\lambda_\star(\mu)}$}

We want to prove that the function $\lambda_\star(\mu)$ defined in
Eq.~\reff{def.lambdastar} is an increasing function of $\mu$ for $\mu \ge 0$,
or in other words that the function
\begin{eqnarray}
   & &
   {d \over d\mu} \, \lambda_\star(\mu)
   \;=\;
   {e^{-2\mu \arctan\mu} \: \mu
    \over
    (1+ 4 \mu^2)^2 \, \sqrt{1+\mu^2}
   }
   \: \times
           \nonumber \\[1mm]
   & & \quad
   \bigg[ 24 \, e^{2\mu \arctan\mu} \sqrt{1+\mu^2}
         \:-\:
         (18 + 24 \mu^2)
         \:-\: 4 (1 + \mu^2) (1 + 4 \mu^2) {\arctan\mu \over \mu}
   \bigg]
   \qquad
 \label{def.lambdastar.deriv}
\end{eqnarray}
is nonnegative for all $\mu \ge 0$.
The proof is unfortunately a bit ugly.

\begin{figure}[t]
\begin{center}
\includegraphics[width=0.8\textwidth]{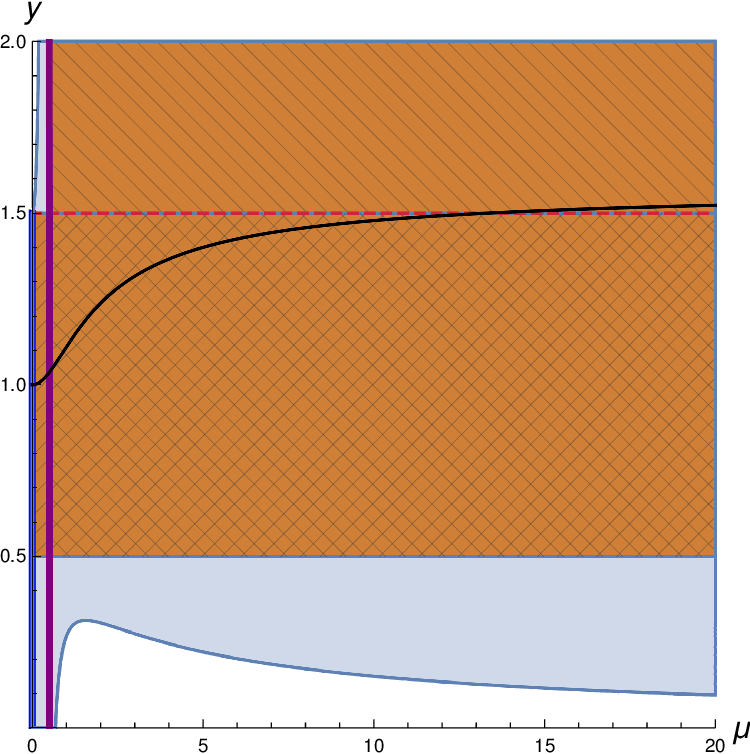}
\end{center}
   \caption{The region in the $(\mu,y)$-plane where $f(\mu,y) \ge 0$
            is shaded in light blue;
            the region where we prove that $f(\mu,y) \ge 0$
            is mesh-shaded in orange.
            The black solid curve is $y = (\arctan\mu) / [\mu/\sqrt{1+\mu^2}]$.
            The dashed red line is $y = 3/2$, which is the upper limit
            of the validity of $f(\mu,y) \ge 0$ for very small $\mu$.
            The thick blue and purple vertical lines indicate places where
            we have a simple proof that $f(\mu,y) \ge 0$.
           }
   \label{fig.muyplot}
\end{figure}

We shall focus on the quantity in square brackets
in Eq.~\reff{def.lambdastar.deriv} and prove that it is nonnegative.
We begin by observing that the function $(\arctan\mu) / [\mu/\sqrt{1+\mu^2}]$
is an increasing function of $\mu$ on the interval $\mu \ge 0$,
which runs from 1 at $\mu = 0$ to $\pi/2$ as $\mu \to +\infty$;
this follows from the fact that
\be
   {d \over d\mu} \: {\arctan\mu \over \mu/\sqrt{1+\mu^2}}
   \;=\;
   {\mu \,-\, \arctan\mu \over \mu^2 \sqrt{1+\mu^2}}
   \;\ge\;
   0
   \;.
\ee
So we write
\be
   \arctan\mu
   \;=\;
   y \: {\mu \over \sqrt{1+\mu^2}}
 \label{def.y}
\ee
and define the function of two variables
\be
   f(\mu,y)
   \;\eqdef\;
   24 \, e^{2y \mu^2/\sqrt{1+\mu^2}} \sqrt{1+\mu^2}
         \:-\: (18 + 24 \mu^2)
         \:-\: 4 \sqrt{1 + \mu^2} \, (1 + 4 \mu^2) \, y
   \;,
\ee
in which the arctangent no longer appears.
We need to prove that $f(\mu,y) \ge 0$
on the curve $y = (\arctan\mu) / [\mu/\sqrt{1+\mu^2}]$,
but we will actually prove it in a much larger region of the $(\mu,y)$-plane
--- not quite the whole region where it actually holds,
but a fairly large chunk of it (see Fig.~\ref{fig.muyplot}).

\medskip

{\bf Step 1.}
We have
\be
   f(0,y)  \;=\;  6 - 4y
   \;,
 \label{eq.f.mu=0}
\ee
which implies
\be
   f(0,y)
   \;\ge\;
   0
   \quad\hbox{for $0 \le y \le 3/2$}
 \label{eq.f_ge_0.mu=0}
\ee
(indicated by a thick blue vertical line in Fig.~\ref{fig.muyplot}).

\medskip

{\bf Step 2.}
Using $\exp(x) \ge 1+x$, we have
\be
   f(\mu,y)
   \;\ge\;
   \widehat{f}(\mu,y)
   \;\eqdef\;
   24 \, \big( \sqrt{1+\mu^2} + 2y \mu^2 \big)
         \:-\: (18 + 24 \mu^2)
         \:-\: 4 \sqrt{1 + \mu^2} \, (1 + 4 \mu^2) \, y
   \;.
   \quad
 \label{def.fLB}
\ee
The function $\widehat{f}(\mu,y)$ is of the form $a+by$,
and the coefficients $a$ and $b$ are both nonnegative
when $0.371412 \ltapprox \mu \ltapprox 0.676097$.
In particular, for $\mu = 1/2$ we have
$\widehat{f}(1/2,y) = (12\sqrt{5}-24) + (12 - 4\sqrt{5}) y$.
Therefore
\be
   f(1/2,y)
   \;\ge\;
   \widehat{f}(1/2,y)
   \;\ge\;
   0
   \quad\hbox{for all $y \ge 0$}
 \label{eq.f.mu=1/2}
\ee
(indicated by a thick purple vertical line in Fig.~\ref{fig.muyplot}).

\medskip

{\bf Step 3.}
We now work on the derivative $\partial f/\partial \mu$, which is
\be
   {\partial f \over \partial\mu} (\mu,y)
   \;=\;
   12 \mu
   \biggl[
   {2 \, e^{2y \mu^2/\sqrt{1+\mu^2}} \,
      \big[ \sqrt{1+\mu^2} \,+\, (4+2\mu^2)y \big]
    \over
    1 + \mu^2
   }
   \:-\: 4
   \:-\: {3 + 4\mu^2 \over \sqrt{1+\mu^2}} \, y
   \biggr]
   \;.
 \label{eq.f.deriv}
\ee
Using $\exp(x) \ge 1+x$, this gives
\be
   {\partial f \over \partial\mu} (\mu,y)
   \;\ge\;
   g(\mu,y)
   \;\eqdef\;
   12 \mu
   \biggl[
   {2 \, \Big( 1 + {2y \nu \over \sqrt{1+\mu^2}} \Big) \,
      \big[ \sqrt{1+\mu^2} \,+\, (4+2\mu^2)y \big]
    \over
    1 + \mu^2
   }
   \:-\: 4
   \:-\: {3 + 4\mu^2 \over \sqrt{1+\mu^2}} \, y
   \biggr]
   \;.
 \label{def.g}
\ee
The function $g(\mu,y)$ is now a quadratic in $y$,
and it is not difficult to prove that
\be
   (\partial f/\partial \mu)(\mu,y)
   \;\ge\;
   g(\mu,y)
   \;\ge\;
   0
   \quad\hbox{for all $\mu \ge 0$ and $y \ge 1/2$}
      \;.
   \label{eq.gnuy}
\ee
Indeed, if we make the substitution $y = \smhalf + \widetilde{y}$, we have
\be
   g(\mu,y)
   \;=\;
   {6 \mu \over (1+\mu^2)^{3/2}}
   \:
   \Bigl[ (1 + 5\mu^2)
             \:+\: \big( -6 + 16 (1+\mu^2)^{3/2} + 18\mu^2 + 8\mu^4 \big)
                           \widetilde{y}
             \:+\: (32 \mu^2 + 16\mu^4) \widetilde{y}^2
   \Bigr]
   \;,
\ee
in which all three coefficients are manifestly nonnegative for $\mu \ge 0$.

\medskip

{\bf Conclusion of the argument.}
Combining Eq.~\reff{eq.f_ge_0.mu=0} with Eq.~\reff{eq.gnuy}, we conclude that
\be
   f(\mu,y) \;\ge\; 0
   \quad\hbox{for $\mu \ge 0,\: 1/2 \le y \le 3/2$}
\ee
(NE--SW shaded region lying below the dashed red line
 in Fig.~\ref{fig.muyplot}).
In particular, the part $0 \le \mu \ltapprox 13.3057$
of the curve $y = (\arctan\mu) / [\mu/\sqrt{1+\mu^2}]$
---
that is, the part of the black curve lying below the dashed red line
 in Fig.~\ref{fig.muyplot} ---
is contained in this region.

Similarly, combining Eq.~\reff{eq.f.mu=1/2} with Eq.~\reff{eq.gnuy},
we conclude that
\be
   f(\mu,y) \;\ge\; 0
   \quad\hbox{for $\mu \ge 1/2,\: y \ge 1/2$}
\ee
(NW--SE shaded region in Fig.~\ref{fig.muyplot}).
In particular, the part $\mu \ge 1/2$
of the curve $y = (\arctan\mu) / [\mu/\sqrt{1+\mu^2}]$
is contained in this latter region.

These two regions together cover the whole
curve $y = (\arctan\mu) / [\mu/\sqrt{1+\mu^2}]$,
thereby completing the proof that $\lambda_\star(\mu)$
is an increasing function of $\mu$.


\subsection{Behavior of the function
  $\bm{\Lambda(\theta) = \Lambda(\theta;\mu,\lambda)}$}

We shall study the behavior of the function
$\Lambda(\theta) = \Lambda(\theta;\mu,\lambda)$ 
defined by Eq.~\reff{eq.soln.Lambdatheta};
we~always assume that $\mu > 0$, $0 < \lambda < 1$
and $0 \le \theta \le \pi/2$.
In what follows, $\mu > 0$ is always fixed;
only $\lambda$ and $\theta$ are variable.

{}From Eq.~\reff{eq.soln.Lambdatheta} we have
\be
   \Lambda(\theta)
   \;=\;
   \cos\theta  \:+\: (\lambda - 1) e^{2\mu\theta}
     \:+\:
   {3 (e^{2\mu\theta} - \cos\theta - 2\mu\sin\theta)  \over  1 + 4\mu^2}
   \;,
\ee
which takes the value $\lambda$ at $\theta = 0$.
Its derivative with respect to $\theta$ is
\be
   \Lambda'(\theta)
   \;=\;
   -\sin\theta  \:+\:  2\mu (\lambda - 1) e^{2\mu\theta} 
        \:+\:
   {3 (2\mu e^{2\mu\theta} + \sin\theta - 2\mu\cos\theta)  \over  1 + 4\mu^2}
   \;,
\ee
which takes the value $2\mu (\lambda-1) < 0$ at $\theta = 0$.
Clearly, both $\Lambda$ and $\Lambda'$ are strictly increasing functions
of $\lambda$ at fixed $\mu,\theta$.

Observe now that $\Lambda'(\theta)$ vanishes when (and only when)
$\lambda$ takes the special value
\be
   \lambda^\natural(\mu,\theta)
   \;\eqdef\;
   {e^{-2\mu\theta} [3\mu\cos\theta + (2\mu^2-1)\sin\theta]
       \,+\, 4\mu^3 - 2\mu
    \over
    \mu (1+4\mu^2)
   }
   \;.
\ee
In fact, when $\lambda = \lambda^\natural(\mu,\theta)$
we have $\Lambda'(\theta) = 0$ (by construction) and
the amazingly simple value
\be
   \left. \Lambda(\theta;\mu,\lambda) \right|
                    _{\lambda = \lambda^\natural(\mu,\theta)}
   \;\,=\;\,
   \cos\theta \,-\, {\sin\theta \over \mu}
   \;.
\ee
It follows that whenever $\lambda < \lambda^\natural(\mu,\theta)$,
we have
\be
   \Lambda'(\theta) \:<\: 0
   \;\hbox{and}\;
   \Lambda(\theta) \:<\: \cos\theta \,-\, {\sin\theta \over \mu}
   \;;
 \label{eq.concl1}
\ee
and whenever $\lambda > \lambda^\natural(\mu,\theta)$,
we have
\be
   \Lambda'(\theta) \:>\: 0
   \;\hbox{and}\;
   \Lambda(\theta) \:>\: \cos\theta \,-\, {\sin\theta \over \mu}
   \;.
 \label{eq.concl2}
\ee

\begin{figure}[t]
\begin{center}
\includegraphics[width=0.6\textwidth]{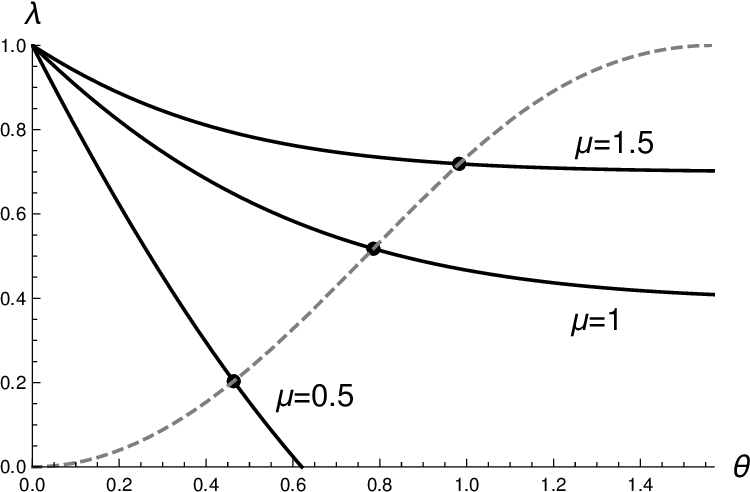}
\end{center}
   \caption{$\lambda^\natural(\mu,\theta)$ as a function of $\theta$
               for $\mu = 0.5$, 1, 1.5.
            The dashed curve is defined parametrically by
            $\theta = \arctan\mu$ and $\lambda = \lambda_\star(\mu)$.
           }
   \label{fig.lambdanatural}
\end{figure}

On the other hand,
\be
   {\partial \lambda^\natural(\mu,\theta)
    \over
    \partial \theta
   }
   \;=\;
   - \, e^{-2\mu\theta} \,
            \Bigl( \sin\theta \,+\, {\cos\theta  \over  \mu} \Bigr)
   \;<\;
   0
   \;.
\ee
It follows that $\lambda^\natural(\mu,\theta)$
is a strictly decreasing function of $\theta$
throughout the interval $0 \le \theta \le \pi/2$.
Moreover, $\lambda^\natural(\mu,\theta)$ takes the value 1 at $\theta=0$
and decreases to $\lambda_\star(\mu)$ [defined in Eq.~\reff{def.lambdastar}]
at $\theta = \arctan \mu$ (see Fig.~\ref{fig.lambdanatural}).
Therefore,
\be
   \lambda^\natural(\mu,\theta) \,\ge\, \lambda_\star(\mu)
      \quad \hbox{whenever $0 \le \theta \le \arctan \mu$}
   \;,
 \label{ineq.lambnatural.lambdastar}
\ee
with strict inequality except at the endpoint $\theta =  \arctan \mu$.

Since $\lambda^\natural(\mu,\theta)$
is a strictly decreasing function of $\theta$,
we can also define the inverse function $\theta^\natural(\mu,\lambda)$;
it is a strictly decreasing function of $\lambda$.
This function is well-defined on the interval
$\lambda^\natural(\mu,\pi/2) \le \lambda \le 1$,
but we shall use it only on the smaller interval
$\lambda_\star(\mu) \le \lambda \le 1$.
We observe that $\lambda < \lambda^\natural(\mu,\theta)$
if and only if $\theta < \theta^\natural(\mu,\lambda)$;
this corresponds to the point $(\theta,\lambda)$
lying below the solid curve in Fig.~\ref{fig.lambdanatural}.
Similarly, $\lambda > \lambda^\natural(\mu,\theta)$
if and only if $\theta > \theta^\natural(\mu,\lambda)$;
this corresponds to the point $(\theta,\lambda)$
lying above the solid curve in Fig.~\ref{fig.lambdanatural}.

\bigskip

{\bf Case $\bm{\lambda < \lambda_\star(\mu)}$.}
The hypotheses $\lambda < \lambda_\star(\mu)$
and $0 \le \theta \le \arctan \mu$
together imply $\lambda < \lambda^\natural(\mu,\theta)$
[by Eq.~\reff{ineq.lambnatural.lambdastar}],
and hence, by Eq.~\reff{eq.concl1}, $\Lambda'(\theta) < 0$
and $\Lambda(\theta) < \cos\theta$.
So, when $\lambda < \lambda_\star(\mu)$,
the function $\Lambda(\theta)$ is strictly decreasing
on the interval $0 \le \theta \le \arctan \mu$.
Moreover, at the point $\theta = \arctan\mu$
we have, again by Eq.~\reff{eq.concl1},
\be
   \Lambda(\theta) \;<\; \cos\theta \,-\, {\sin\theta \over \mu}
                   \;=\; 0
   \;.
\ee
It follows that $\Lambda(\theta)$ must have a unique zero
in the interval $0 < \theta < \arctan\mu$,
and that $\Lambda'(\theta) < 0$ at this point.

This proves the claim that when $0 < \lambda < \lambda_\star(\mu)$,
the skier halts at some angle $\theta_{\rm halt}(\mu,\lambda)$
in the interval $(0,\arctan\mu)$.
(Since $\Lambda(\theta) < \cos\theta$ for $0 \le \theta \le \arctan \mu$,
 the skier cannot have flown off earlier.)
Moreover, because $\Lambda(\theta)$ crosses zero with a nonzero slope,
the singularity at $\theta = \theta_{\rm halt}(\mu,\lambda)$
in the integral \reff{eq.time} is integrable,
and the skier halts after a finite time
\be
   T_{\rm halt}
   \;=\;
   \int\limits_0^{\theta_{\rm halt}(\mu,\lambda)}
   \!\!\!\!
      {d\theta' \over \Bigl[ \displaystyle {g \over R} \,
                             \Lambda(\theta') \Bigr]^{1/2}}
   \;<\; \infty
   \;.
\ee
Finally, $\theta_{\rm halt}(\mu,\lambda)$
is an increasing function of $\lambda$
because, on the relevant interval,
$\Lambda(\theta;\mu,\lambda)$ is an increasing function of $\lambda$
and a decreasing function of $\theta$.

This behavior is illustrated in the curves $\lambda < \lambda_\star$
of Fig.~\ref{fig.Lambda.theta}.

\bigskip

{\bf Case $\bm{\lambda = \lambda_\star(\mu)}$.}
When $\lambda = \lambda_\star(\mu)$,
the foregoing argument shows that $\Lambda'(\theta) < 0$
for $0 \le \theta < \arctan\mu$;
and of course $\Lambda(\theta) = \Lambda'(\theta) = 0$
at $\theta = \arctan\mu$.
Therefore $\Lambda(\theta) > 0$ for $0 \le \theta < \arctan\mu$,
and $\theta_{\rm halt} = \arctan\mu$.

Since $\Lambda(\theta) = \Lambda'(\theta) = 0$ at $\theta = \arctan\mu$,
the singularity at $\theta = \arctan\mu$
in the integral \reff{eq.time} is nonintegrable,
and the skier comes to rest at $\theta=\arctan\mu$
asymptotically as $t \to +\infty$.

This behavior is illustrated in the curve $\lambda = \lambda_\star$
of Fig.~\ref{fig.Lambda.theta}.

\bigskip

{\bf Case $\bm{\lambda > \lambda_\star(\mu)}$.}
We have just seen that, for $\lambda = \lambda_\star(\mu)$
and $0 \le \theta \le \arctan\mu$,
we have $\Lambda'(\theta) \le 0$ and $\Lambda(\theta) \ge 0$,
with equality at $\theta=\arctan\mu$.
On the other hand, setting $\lambda = \lambda_\star(\mu)$
and $\theta = \arctan\mu +\psi$, we have
\be
   \Lambda'(\arctan\mu + \psi; \mu, \lambda_\star(\mu))
   \;=\;
   {2 \sqrt{1+\mu^2} \over 1 + 4\mu^2}
      \: \bigl[ 2\mu e^{2\mu\psi} \,+\, \sin\psi \,-\, 2\mu \cos\psi \bigr]
   \;,
 \label{eq.Lambdaprime.psi}
\ee
which is easily seen to be $\ge 0$ for $0 \le \psi \le \pi$
(since $e^{2\mu\psi} \ge 1$, $\sin\psi \ge 0$ and $\cos\psi \le 1$).
It follows that, when $\lambda = \lambda_\star(\mu)$,
we have $\Lambda'(\theta) \ge 0$ and $\Lambda(\theta) \ge 0$
for $\arctan\mu \le \theta \le \pi/2$,
and hence $\Lambda(\theta) \ge 0$
throughout the interval $0 \le \theta \le \pi/2$.
And since $\Lambda(\theta;\mu,\lambda)$ is a strictly increasing
function of $\lambda$ for fixed $\mu,\theta$,
we have $\Lambda(\theta) > 0$ for $0 \le \theta \le \pi/2$
whenever $\lambda > \lambda_\star(\mu)$.
This proves that for $\lambda > \lambda_\star(\mu)$
the skier cannot halt.

Now fix $\lambda > \lambda_\star(\mu)$.
For $0 \le \theta \le \theta^\natural(\mu,\lambda)$ we have
$\Lambda(\theta) \le \cos\theta - (\sin\theta)/\mu < \cos\theta$
[by Eq.~\reff{eq.concl1}];
for $\theta^\natural(\mu,\lambda) < \theta \le \pi/2$,
the function $\Lambda(\theta)$ is strictly increasing [by Eq.~\reff{eq.concl2}]
while $\cos\theta$ is strictly decreasing;
and at $\theta = \pi/2$ we have $\Lambda(\theta) > 0 = \cos\theta$.
It follows that the equation $\Lambda(\theta) = \cos\theta$
has a unique solution $\theta_{\rm fly}(\mu,\lambda)$
in the interval $[0,\pi/2]$, and this solution satisfies
$\theta^\natural(\mu,\lambda) < \theta_{\rm fly}(\mu,\lambda) < \pi/2$.
Moreover, $\theta_{\rm fly}(\mu,\lambda)$
is a decreasing function of $\lambda$,
because $\Lambda(\theta;\mu,\lambda) - \cos\theta$
is an increasing function of both $\theta$ and $\lambda$
in the relevant interval.

This behavior is illustrated in the curves $\lambda > \lambda_\star$
of Fig.~\ref{fig.Lambda.theta}.
We conjecture that
$\theta_{\rm fly}(\mu,\lambda)$ is an increasing function of $\mu$
at each fixed $\lambda$, but we do not have a proof.

%
%
%
%
%
%

\section{Proofs for the loop-the-loop}

\subsection{Behavior of the function $\bm{\lambdatilde(\pi/2,\mu)}$}

We want to prove that $\lambdatilde(\pi/2,\mu)$,
which forms the boundary between the
``halts in the first quadrant'' and ``flies off the second quadrant'' phases,
is an increasing function of $\mu$.
From Eq.~\reff{def.lambdatilde} we obtain
\be
   {d \over d\mu} \, \lambdatilde(\pi/2,\mu)
   \;=\;
   {-24\mu \:+\: 6 e^{\pi\mu} [1 - 4 \mu^2 + \pi\mu (1+ 4\mu^2)]
    \over
    (1+4\mu^2)^2
   }
   \;.
 \label{eq.B1.a}
\ee
We have $1 - 4\mu + 4\mu^2 = (1 - 2\mu)^2 \ge 0$
and hence $1 + 4\mu^2 \ge 4\mu$
(this is just the arithmetic-geometric mean inequality).
So the term in square brackets in Eq.~\reff{eq.B1.a} is
$\ge 1 + (4\pi-4) \mu^2 \ge 0$.
We can therefore use the lower bound $e^{\pi\mu} \ge 1 + \pi\mu$
to deduce
\begin{subeqnarray}
   6 e^{\pi\mu} [1 - 4 \mu^2 + \pi\mu (1+ 4\mu^2)] \,-\, 24\mu
   & \ge &
   6 \, (1 + \pi\mu) \, [1 + (4\pi-4) \mu^2] \,-\, 24\mu
         \\[2mm]
   & = &
   6 \bigl[ 1 - (4-\pi)\mu + (4\pi-4)\mu^2 + (4\pi^2 - 4\pi) \mu^3 \bigr]
      \nonumber \\
   \;.
\end{subeqnarray}
But the quadratic $1 \,-\, (4-\pi)\mu \,+\, (4\pi-4)\mu^2$
is everywhere positive,
so the numerator of Eq.~\reff{eq.B1.a} is positive, and we are done.

\subsection{Behavior of the function $\bm{\lambdahat(\pi,\mu)}$}

We want to prove that $\lambdahat(\pi,\mu)$,
which forms the boundary between the
``flies off the second quadrant'' phase
and the two upper phases in Fig.~\ref{fig.loop-the-loop_phasediag},
is an increasing function of $\mu$.
{}From Eq.~\reff{def.lambdahat} we obtain
\be
   {d \over d\mu} \, \lambdahat(\pi,\mu)
   \;=\;
   {-24\mu \:+\: 6\pi e^{2\pi\mu} [1 \,-\, (4/\pi)\mu \,+\, 4\mu^2]
    \over
    (1+4\mu^2)^2
   }
   \;.
 \label{eq.B2.a}
\ee
By reasoning similar to that in the previous subsection,
we show that the numerator of Eq.~\reff{eq.B2.a} is positive.

\subsection{Behavior of the function $\bm{\lambdatilde(2\pi,\mu)}$}

We now consider the function $\lambdatilde(2\pi,\mu)$,
which is given by Eq.~\reff{def.lambdatilde.2pi}.
It is negative for $0 < \mu < 1/\sqrt{2}$
and positive for $\mu > 1/\sqrt{2}$.
We wish to prove that it is also increasing when $\mu > 1/\sqrt{2}$.
But this is easy:
the function $e^{4\pi\mu} - 1$ is positive and increasing when $\mu > 0$;
and the function
\be
   {4\mu^2 - 2 \over 1 + 4\mu^2}
   \;=\;
   1 \:-\: {3 \over 1 + 4\mu^2}
\ee
is positive and increasing when $\mu > 1/\sqrt{2}$.
So their product is positive and increasing when $\mu > 1/\sqrt{2}$.


As will be shown in the next subsection,
the function $\lambdatilde(2\pi,\mu)$
forms the boundary between the
``flies off the second quadrant''
and ``halts in the fourth quadrant'' phases
when $\mu \ge \mu_{\rm crit} \approx 0.713089$.
So the proof given here for $\mu > 1/\sqrt{2} \approx 0.707107$
is sufficient to handle this region.

\subsection{Uniqueness of $\bm{\mu_{\rm crit}}$}

We wish to prove that there is a
unique value $\mu_{\rm crit} \approx 0.713089$
such that the function
\begin{subeqnarray}
   f(\mu)
   & \eqdef &
   \lambdahat(\pi,\mu) \,-\, \lambdatilde(2\pi,\mu)
          \\[2mm]
   & = &
   {e^{2\pi\mu} \over 1 + 4\mu^2} \:
       \bigl[3 \,+\, (2 - 4\mu^2) e^{2\pi\mu} \bigr]
\end{subeqnarray}
is positive for $0 \le \mu < \mu_{\rm crit}$,
zero for $\mu = \mu_{\rm crit}$,
and negative for $\mu > \mu_{\rm crit}$.
We remove the positive prefactor and concentrate on
\be
   g(\mu)  \;=\; 3 \,+\, (2 - 4\mu^2) e^{2\pi\mu}
   \;.
\ee
We have $g(0) = 5$ and
\be
   g'(\mu)  \;=\; 4 e^{2\pi\mu} \bigl[ \pi \,-\, 2\mu \,-\, 2\pi \mu^2 \bigr]
   \;.
\ee
This is a quadratic that is positive for
$0 \le \mu < (\sqrt{1+2\pi^2} - 1)/(2\pi) \approx 0.565642$
and negative for larger $\mu$.
So $g(\mu)$ is positive and increasing for $0 \le \mu \ltapprox 0.565642$,
and decreasing thereafter.
Since $\lim\limits_{\mu\to +\infty} g(\mu) = -\infty$,
the function $g$ clearly has a unique root,
after which it is negative.


\addcontentsline{toc}{section}{References}
\begin{thebibliography}{99}

\bibitem{mechanics_books}
See e.g.\ D. Kleppner and R. Kolenkow,
   {\em An Introduction to Mechanics}\/, 2nd ed.\ 
   (Cambridge University Press, Cambridge, 2014),
Problems~5.1 (loop-the-loop) and 5.6 (block sliding down a sphere);
D. Morin, {\em Introduction to Classical Mechanics}\/
   (Cambridge University Press, New York, 2008),
Exercises~5.39 (loop-the-loop) and
5.53 (skier on a frictionless hemisphere of finite mass~$M$,
 which is considerably more difficult than the usual case $M=\infty$).

\bibitem{Franklin_80}   L.P. Franklin and P.I. Kimmel,
   ``Dynamics of circular motion with friction,''
   Amer. J. Phys. {\bf 48}, 207--210 (1980).

\bibitem{Mania_02}  A.J. Mania, A.W. Mol and C.S.S. Brand\~ao,
   ``Sliding block on a semicircular track with friction,''
   Revista Brasileira de Ensino de F\'{\i}sica {\bf 24}, 312--316 (2002).

\bibitem{Mungan_03}  C.E. Mungan, ``Sliding on the surface of a rough sphere,''
   Phys. Teacher {\bf 41}, 326--328 (2003).

\bibitem{Hite_04}  G.E. Hite, ``The sled race,''
   Amer. J. Phys. {\bf 72}, 1055--1058 (2004).

\bibitem{Prior_07}  T. Prior and E.J. Mele,
   ``A block slipping on a sphere with friction: Exact and perturbative
   solutions,''
   Amer. J. Phys. {\bf 75}, 423--426 (2007).

\bibitem{DeLange_08}  O.L. de Lange, J. Pierrus, T. Prior and E.J. Mele,
   ``Comment on `A block slipping on a sphere with friction:
   Exact and perturbative solutions',''
   Amer. J. Phys. {\bf 76}, 92--93 (2008).

\bibitem{Klobus_11}  W. K\l{}obus, ``Motion on a vertical loop with friction,''
   Amer. J. Phys. {\bf 79}, 913--918 (2011).

\bibitem{Nahin_15}  P.J. Nahin, {\em Inside Interesting Integrals}\/
   (Springer, New York, 2015), pp.~112--114.

\bibitem{Gonzalez-Cataldo_17}  F. Gonz\'alez-Cataldo, G. Guti\'errez
   and J.M. Y\'a\~nez,
   ``Sliding down an arbitrary curve in the presence of friction,''
   Amer. J. Phys. {\bf 85}, 108--114 (2017).
   Extended version available at \url{https://arxiv.org/abs/1512.00515}

\bibitem{DelPino_18}  L.A. del Pino and S. Curilef,
   ``Comment on `Sliding down an arbitrary curve in the presence of friction',''
   Amer. J. Phys. {\bf 86}, 470--471 (2018).
   
\bibitem{F=ma}
   See Ref.~\onlinecite{Mungan_03}, Eqs.~(1) and (3).
   See also Ref.~\onlinecite{Gonzalez-Cataldo_17} for a generalization
   to an arbitrary curve in the vertical plane,
   using the Frenet--Serret formalism.

\bibitem{note_static_friction}
   Taking literally the equations
   \reff{eq.skier.radial}/\reff{eq.skier.tangential},
   the skier would reverse direction
   when $\dot{\theta} = 0$
   and begin climbing back up the hill.
   But this is not, of course, what actually happens.
   Rather, when the skier halts, static friction takes over,
   and the skier remains forever at rest;
   this occurs because static friction is governed by the
   {\em inequality}\/ $|F_s| \le \mu N$,
   not the equality $|F_s| = \mu N$.

\bibitem{alternate_solution}
   See also Ref.~\onlinecite{Mungan_03}, Appendix~A
   for an alternate approach to solving
   Eqs.~\reff{eq.diffeqn.Ntheta} and \reff{eq.diffeqn.Lambdatheta},
   which does not require the student to be familiar with the method
   of integrating factors.
 
\bibitem{speed_as_a_function_of_angle}
   See Ref.~\onlinecite{Mungan_03}, Eq.~(A3);
   Ref.~\onlinecite{Prior_07}, Eq.~(25);
   Ref.~\onlinecite{DeLange_08}, Eq.~(2);
   Ref.~\onlinecite{Gonzalez-Cataldo_17}, Eq.~(23),
   or Eq.~(24) in the arXiv version.

\bibitem{differential_equation}
   See Ref.~\onlinecite{Mungan_03}, Eq.~(5);
   Ref.~\onlinecite{Prior_07}, Eq.~(17);
   and Ref.~\onlinecite{DeLange_08}, Eq.~(1).
   See also Ref.~\onlinecite{Gonzalez-Cataldo_17}, Eq.~(9)
   for a generalization to an arbitrary curve in the vertical plane.

\bibitem{work-energy}
   This approach is taken, for instance,
   in Ref.~\onlinecite{Prior_07}, Eqs.(9) and (17);
   in Ref.~\onlinecite{DeLange_08}, Eq.~(1);
   and in Ref.~\onlinecite{DelPino_18}
   for an arbitrary curve in the vertical plane.

\bibitem{ref_thetastar}
   See Ref.~\onlinecite{DeLange_08};
   Ref.~\onlinecite{Gonzalez-Cataldo_17} (arXiv version), Eq.~(59) ff.

\bibitem{lambdastar_mu}
   See Ref.~\onlinecite{DeLange_08}, Eq.~(5);
   Ref.~\onlinecite{Gonzalez-Cataldo_17} (arXiv version), Eq.~(62).
   This latter paper also gives analogous formulae for the
   parabola, cycloid, catenary and ellipse.

\bibitem{note_equilibrium}
   It can be seen directly from the equations of motion
   \reff{eq.skier.radial}/\reff{eq.skier.tangential}
   that this ``asymptotic equilibrium'' position
   can only be $\theta = \arctan\mu$.
   To see this, observe that $\sgn(\dot{\theta}) = +1$ for all $t \ge 0$,
   and that $\dot{\theta}, \ddot{\theta} \to 0$ as $t \to +\infty$.
   Combining Eqs.~\reff{eq.skier.radial}/\reff{eq.skier.tangential}
   as $t \to +\infty$
   then yields $N = mg \cos\theta$ and $\theta = \arctan\mu$.
   This argument does {\em not}\/ apply to the ``subcritical'' trajectories
   in which the skier halts after a finite time,
   since in these trajectories $\dot{\theta} = 0$ but $\ddot{\theta} \neq 0$
   at the halting time.

\bibitem{phase_diagram}
   See Ref.~\onlinecite{DeLange_08}, Fig.~1;
   Ref.~\onlinecite{Gonzalez-Cataldo_17} (arXiv version), Fig.~13.
   This latter figure also shows the phase diagram for the
   parabola, cycloid, catenary and ellipse.

\bibitem{refs_for_fig.Lambda.theta}
   Compare Ref.~\onlinecite{Mungan_03}, Fig.~2;
   Ref.~\onlinecite{Gonzalez-Cataldo_17}, Fig.~3,
   or Fig.~4 in the arXiv version.
   The latter plot shows different values of $\mu$
   for the same $\lambda = 0.6$,
   which is complementary to our Fig.~\ref{fig.Lambda.theta}.

\bibitem{ref_for_figs_thetahalt+thetafly}
   See Ref.~\onlinecite{DeLange_08}, Fig.~2
   for a superposed version of our
   Figs.~\ref{fig.thetahalt} and \ref{fig.thetafly}
   that highlights this discontinuity.

\bibitem{ref_Maxwell}
J.C. Maxwell, {\em Matter and Motion}\/
   (Society for Promoting Christian Knowledge, London, 1876);
   reprinted by Dover, New York, 1952
   and Cambridge University Press, Cambridge, 2010.
After stating (p.~20) what he sees as
``the general maxim of physical science'' ---
namely, ``the same causes will always produce the same effects'' ---
Maxwell goes on to observe (p.~21) that
\begin{quotation}
   There is another maxim which must not be confounded
   with [this one],
   which asserts ``That like causes produce like effects.''

   This is only true when small variations in the
   initial circumstances produce only small variations
   in the final state of the system.
   In a great many physical phenomena this condition is satisfied;
   but there are other cases in which a small initial variation
   may produce a very great change in the final state of the system,
   as when the displacement of the `points' causes a railway train
   to run into another instead of keeping its proper course.
\end{quotation}

\bibitem{ref_supplementary}  Supplementary Materials are provided at
{\bf URL to be inserted by AIPP}

\bibitem{Ivchenko_21}  V. Ivchenko, ``Sliding down a rough curved hill'',
   European J. Phys. {\bf 42}, 025005 (2021).

\bibitem{Balart_19}  L. Balart and S. Belmar-Herrera,
  ``Particle sliding down an arbitrary concave curve in the
   Lagrangian formalism'',
   Amer. J. Phys. {\bf 87}, 982--985 (2019).

\bibitem{Mejia_20}  G.M. Mej\'{\i}a, J.M. Betancourt, C.D. Forero,
   N. Avil\'an, F.J. Rodr\'{\i}guez, L. Quiroga and N.F. Johnson,
   ``Dynamics of a round object moving along curved surfaces with friction'',
   Amer. J. Phys. {\bf 88}, 229--237 (2020).

\bibitem{loop-the-loop_reverses}
   The qualitative behavior after the particle reverses direction is,
   however, very simple.
   As will be seen below,
   the particle can come instantaneously to rest
   only in the lower half of the loop-the-loop
   ($-\pi/2 \le \theta \le \pi/2$ modulo $2\pi$).
   After this happens, the particle simply oscillates back and forth,
   with constant amplitude if $\mu = 0$
   and with decreasing amplitude if $\mu > 0$.

\bibitem{loop-the-loop_Lambda_theta}
   See Ref.~\onlinecite{Franklin_80}, Eq.~(12);
   Ref.~\onlinecite{Mania_02}, Eq.~(6);
   Ref.~\onlinecite{Klobus_11}, Eq.~(6).
   See also Ref.~\onlinecite{Mania_02} for a generalization
   that includes a viscous drag force $-\beta v^2$.

\end{thebibliography}
\end{document}